\documentclass[twoside,11pt]{article}

\usepackage{jmlr2e}

\usepackage{stackengine}

\setstackgap{S}{8pt}

\usepackage{amsmath,amssymb,comment,tikz,pgf,bm}
\usepackage{amsopn, amsfonts, leftidx}
\usetikzlibrary{shapes}

\usepackage{algorithm}
\usepackage{algpseudocode}
\usepackage{graphicx}
\usepackage{booktabs}
\usepackage{listings}
\usepackage[mode=buildnew]{standalone}

\newenvironment{thma}[1]{\par\noindent{\scshape Theorem #1\ }\em}{\em \\}

\newtheorem{assumption}{Assumption}[section]

\DeclareMathOperator{\dis}{dis}

\DeclareMathOperator{\pa}{pa}
\DeclareMathOperator{\de}{de}

\DeclareMathOperator{\mb}{mb}

\DeclareMathOperator{\E}{\mathbb{E}}

\def\ci{\perp\!\!\!\perp}

\usepackage[normalem]{ulem}
\newcommand{\stkout}[1]{\ifmmode\text{\sout{\ensuremath{#1}}}\else\sout{#1}\fi}

\jmlrheading{1}{2021}{8-1}{8/1}{/}{shpitser21}{Ilya Shpitser, Zach Wood-Doughty and Eric Tchetgen Tchetgen}

\ShortHeadings{The Proximal ID Algorithm}{Shpitser, Wood-Doughty and Tchetgen Tchetgen}
\firstpageno{1}

\begin{document}

\title{
The Proximal ID Algorithm
}

\author{\name Ilya Shpitser \email ilyas@cs.jhu.edu\\
       \addr Department of Computer Science\\
	   Johns Hopkins University\\
       Baltimore, MD 21218, USA
       \AND
       \name Zach Wood-Doughty \email zach@northwestern.edu \\
       \addr Department of Computer Science\\
	   Northwestern University\\
       Evanston, IL 60208, USA
       \AND
       \name Eric J.\ Tchetgen Tchetgen \email ett@wharton.upenn.edu\\
       \addr Department of Statistics\\
       The Wharton School\\
       3620 Locust Walk, Philadelphia, PA 19104, USA
	 }

\editor{Anonymous}

\maketitle

\begin{abstract}
Unobserved confounding is a fundamental obstacle to establishing valid causal conclusions from observational data.
Two complementary types of approaches have been developed to address this obstacle: obtaining identification using fortuitous external aids, such as instrumental variables or proxies, or by means of the ID algorithm, using Markov restrictions on the full data distribution encoded in graphical causal models.
In this paper we aim to develop a synthesis of the former and latter approaches to identification in causal inference to yield the most general identification algorithm in multivariate systems currently known -- the proximal ID algorithm.  In addition to being able to obtain nonparametric identification in all cases where the ID algorithm succeeds, our approach allows us to systematically exploit proxies to adjust for the presence of unobserved confounders that would have otherwise prevented identification.
In addition, we outline a class of estimation strategies for causal parameters identified by our method in an important special case.  We illustrate our approach by simulation studies
and a data application.
\end{abstract}



\section{Introduction}
\label{sec:intro}

Understanding cause effect relationships is a crucial task in empirical science and rational decision making.  An extensive line of work spanning multiple communities has placed modern causal inference on a rigorous footing, using the language of potential outcomes, random variables representing responses to hypothetical interventions, as well as the language of graphical models and structural equations \citep{pearl09causality,thomas13swig}.

Modern causal inference conceptualizes cause effect relationships between treatments and outcomes using hypothetical randomized experiments implemented via an intervention operation \citep{pearl09causality}.  Linking realizations of the observed data distribution with the counterfactual parameters arising in such experiments entails using assumptions encoded in causal models, with the resulting functionals estimated by plug-in estimators or semi-parametric methods \citep{van2003unified,tsiatis06missing}.

As an example, if the causal relationship between a treatment variable $A$ and an outcome $Y$ is masked by 
{non-causal} associations due to a vector of {observed} common causes $\vec{C}$,
then the average causal effect is identified by the covariate adjustment functional and estimated by matching methods, plug-in estimators, or the augmented inverse probability weighted estimator \citep{hernan2010causal,robins94estimation}.

If some of the relevant confounders are unobserved, the situation becomes considerably more difficult.  In general, unobserved confounding prevents identification of causal parameters.  However, a rich literature has been developed that exploits various kinds of causal 
assumptions that allow point identification to be recovered.
A complete characterization of identifiability of a large class of causal parameters in arbitrary causal models with hidden variables has been developed using the language of graphical models, resulting in the ID algorithm and related extensions \citep{tian02on,shpitser06id,shpitser06idc,huang06do,shpitser18medid}.  Celebrated special cases of this approach, such as the front-door model, are able to obtain nonparametric identification in seemingly counter-intuitive situations, such as when a treatment and an outcome share an arbitrarily complicated unobserved common cause.

An extensive complementary line of work on point identification in the presence of unobserved confounding \citep{wright1928tariff,instrumental01angrist,kuroki14measurement} is based on taking advantage of fortuitous external aids (such as the presence of an instrumental variable or other proxy), along with additional assumptions to ensure identification.
A recent line of work of \emph{proximal causal inference} \citep{miao2018identifying,miao2018confounding,tchetgen2020introduction} provides a novel approach to using proxies to deal with unobserved confounding without relying on stringent parametric assumptions.

Recent work developed semiparametric estimators for causal parameters identified within this framework \citep{semiparametric2020cui},
and extended proximal causal inference from point exposure to time-varying treatment settings \citep{tchetgen2020introduction,ying2021proximal,proxy2021deaner},
heterogeneous treatment effect estimation \citep{singh2020kernel},
treatment policies \citep{kallus2021causal},
and mediation analysis \citep{dukes2021proximal,ghassami21proximal}.

Other methods that aim to obtain identification for parameters in hidden variable graphical models include methods that exploit algebraic results \citep{kruskal1976more,kruskal1977three} for models with hidden variable state-spaces \citep{allman2009identifiability,allman2015parameter}, and results on a relationship between proxies and hidden common causes \citep{kuroki14measurement}.  In addition,
identifiability results in models with exactly one hidden variable have been obtained \citep{sharaf2014identifiability}.




In this paper, we aim to unify graphical and proxy-based approaches by providing an approach to identification of counterfactual parameters in the presence of unobserved confounding, called the \emph{proximal ID algorithm} that inherits the advantages of proximal {causal} inference and nonparametric identification via graphical models.  Like the latter approach, the proximal ID algorithm is able to obtain point identification in multivariate structured systems with arbitrarily complex patterns of unobserved confounding, greatly extending the sets of cases where proximal inference may be used.  At the same time, the proximal ID algorithm is able to exploit the presence of fortuitous proxies 
to obtain identification in cases where the ID algorithm would otherwise fail.

Our paper is organized as follows.  In \S \ref{sec:proximal}, we review the standard setting of proximal causal inference, largely following \citep{miao2018identifying,tchetgen2020introduction}.
In \S \ref{sec:proximal-2} we describe a nontrivial extension of this approach to settings where point identification may be obtained, despite the fact that the standard assumptions that proximal causal inference relies on fail.  In \S \ref{sec:id}, we describe the general theory of nonparametric identification in graphical models and the ID algorithm as a prelude to the description of our main contribution, the proximal ID algorithm,
in \S \ref{sec:new}.
\S \ref{sec:dtr} shows how identification theory developed in this paper may be used for identification of  responses to treatment variables being counterfactually set according to a policy.
\S \ref{sec:sims} describes how existing statistical inference methods developed in the proximal causal inference literature \citep{miao2018confounding,tchetgen2020introduction} generalize to the example described in \S \ref{sec:proximal-2}, and illustrates these methods via simulations.  A data application is described in \S \ref{sec:analysis}.  Our conclusions are contained in \S \ref{sec:conclusions}.

\section{Notation and Illustration of Proximal Causal Inference}
\label{sec:proximal}

A standard setting in causal inference assumes the observed data realizations are {independent, identically distributed} (i.i.d.) samples from a distribution on a set of variables $\vec{C},A,Y$, where $A$ is a treatment or exposure of interest, $Y$ is an outcome of interest, and $\vec{C}$ are a set of baseline covariates.  Cause effect relationships are conceptualized by means {of} potential outcome random variables.  As an example, $Y(a)$ denotes the outcome $Y$ had, possibly contrary to fact, treatment $A$ been administered at value $a$.  Potential outcomes are used to define counterfactual parameters, such as the population average causal effect (ACE): $\beta \equiv \E[Y(a)] - \E[Y(a')]$, where $a$ represents the active treatment value, and $a'$ the placebo or control treatment value.  The goal of causal inference is to identify causal parameters such as $\beta$ from the observed data distribution, such as $p(\vec{C},A,Y)$, using causal assumptions that link counterfactual and observed data, and estimate the resulting identifying functional as efficiently and reliably as possible.

A standard assumption in causal inference states that observed outcomes equal counterfactual outcomes had treatments been set to their observed values.  This assumption is known as \emph{consistency} and is often written concisely as $Y = Y(A)$, almost surely.
Aside from consistency, additional assumptions are needed to identify counterfactual parameters.  A popular causal model that yields identification of the population average causal effect is the conditionally ignorable or ``backdoor'' model.  This model makes two crucial assumptions.  The first is that treatment assignment and potential outcomes are independent given baseline covariates, that is $(Y(a) \ci A \mid \vec{C})$ for all values $a$, and the second is that all treatment values have support conditional on $\vec{C}$: $p(A=a \mid \vec{C}) > 0$, almost surely for all $a$.  {Both assumptions must hold for all values $a$ that are of interest in the problem.}  Given these assumptions, as well as consistency, $\beta$ is identified from $p(\vec{C},A,Y)$ by the adjustment functional: $\E[\E[Y \mid a, \vec{C}] - \E[Y \mid a',\vec{C}]]$, which is a special case of the \emph{g-formula} \citep{robins86new}.

\begin{figure}[!t]
\centering
\begin{tikzpicture}[>=stealth, node distance=1.2cm]
    \tikzstyle{format} = [draw, very thick, ellipse,
                          minimum size=0.2cm, inner sep=1pt]
    \tikzstyle{unobs} = [draw, very thick, red, ellipse,
                         minimum size=0.2cm, inner sep=1pt]
    \tikzstyle{square} = [draw, very thick, rectangle,
                         minimum size=0.2cm, inner sep=2pt]

    \begin{scope}
        \path[->, very thick]
            node[format] (A) {$A$}
            node[format, above right of=A] (C) {${C}$}
            node[format, below right of=C] (Y) {$Y$}

		(C) edge[blue] (A)
		(C) edge[blue] (Y)
		(A) edge[blue] (Y)

            node[below of=C, yshift=-0.1cm] (l) {$(a)$}
        ;
    \end{scope}

    \begin{scope}[xshift=3.0cm]
        \path[->, very thick]
            node[format] (A) {$A$}
            node[ above right of=A] (d) {}
            node[format, below right of=d] (Y) {$Y$}

		node[format, xshift=0.6cm, red, left of=d] (U) {${U}$}
		node[format, xshift=-0.6cm, right of=d] (C) {${C}$}

		(C) edge[blue] (A)
		(C) edge[blue] (Y)
		(U) edge[red] (A)
		(U) edge[red] (Y)
		(C) edge[red] (U)
		
		(A) edge[blue] (Y)

            node[below of=d, yshift=-0.1cm] (l) {$(b)$}
        ;
    \end{scope}

    \begin{scope}[xshift=6.0cm]
        \path[->, very thick]
            node[format] (A) {$A$}
            node[ above right of=A] (d) {}
            node[format, below right of=d] (Y) {$Y$}

		node[format, xshift=0.6cm, red, left of=d] (U) {${U}$}
		node[format, xshift=-0.6cm, right of=d] (C) {${C}$}

		node[format, xshift=0.3cm, left of=U] (Z) {$Z$}
		node[format, xshift=-0.3cm, right of=C] (W) {$W$}

		(C) edge[blue] (A)
		(C) edge[blue] (Y)
		(A) edge[blue] (Y)
		(U) edge[red] (A)
		(U) edge[red] (Y)
		
		(A) edge[blue] (Z)
		(W) edge[blue] (Y)
		(C) edge[blue, bend right] (Z)
		(C) edge[blue] (W)
		
		(C) edge[red] (U)
		(U) edge[red, bend left] (W)
		(U) edge[red] (Z)

            node[below of=d, yshift=-0.1cm] (l) {$(c)$}
        ;
    \end{scope}
  
      \begin{scope}[xshift=10.0cm]
        \path[->, very thick]
            node[format] (A) {$A$}
            node[ above right of=A] (d) {}
            node[format, below right of=d] (Y) {$Y$}

		node[format, xshift=0.6cm, red, left of=d] (U) {${U}$}
		node[format, xshift=-0.6cm, right of=d] (C) {${C}$}

		node[format, xshift=0.3cm, left of=U] (Z) {$Z$}
		node[format, xshift=-0.3cm, right of=C] (W) {$W$}

		(C) edge[blue] (A)
		(C) edge[blue] (Y)
		(A) edge[blue] (Y)
		(U) edge[red] (A)
		(U) edge[red] (Y)
		
		(Z) edge[blue] (A)
		(W) edge[blue] (Y)
		(C) edge[blue, bend right] (Z)
		(C) edge[blue] (W)
		
		(C) edge[red] (U)
		(U) edge[red, bend left] (W)
		(U) edge[red] (Z)

            node[below of=d, yshift=-0.1cm] (l) {$(d)$}
        ;
    \end{scope}
  
\end{tikzpicture}
\caption{
(a) A causal diagram corresponding to the conditionally ignorable model.
(b) A causal diagram corresponding to a model where conditional ignorability fails due to the presence of unobserved confounders.
(c) A causal diagram corresponding to a model with unobserved confounding where identification of the 
ACE is possible via proximal causal inference.
{In this model, the control proxy variable $Z$ is caused by the treatment.}
{(d) A causal diagram where identification of the ACE is possible via proximal causal inference, via the same approach as in the model in (c), but where the control proxy variable $Z$ causes the treatment $A$.}
}
\label{fig:backdoor}
\end{figure}
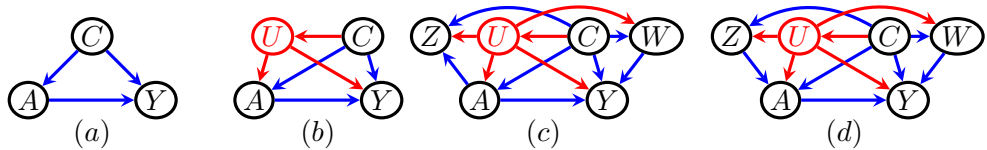

Algebraic restrictions in causal models, such as conditional ignorability, are often displayed visually by means of causal diagrams, particularly directed acyclic graphs (DAGs).
The DAG representing the conditionally ignorable model
is shown in Fig.~\ref{fig:backdoor} (a).   This model implies that $(\vec{C} \ci A(\vec{c}_1) \ci Y(a,\vec{c}_2))$ for any values $a \in {\mathfrak X}_A,\vec{c}_1,\vec{c}_2 \in {\mathfrak X}_{\vec{C}}$, and this in turn implies $(Y(a,\vec{C}) \ci A(\vec{C}) \mid \vec{C})$ which is equivalent to $(Y(a) \ci A \mid \vec{C})$. 
This model may correspond to an observational study in healthcare settings, where one of two treatment alternatives $a=1$ or $a'=0$ are assigned to patients based on their baseline characteristics $\vec{C}$, in hopes of improving their outcome $Y$.  The dependence of $A$ on $\vec{C}$ would represent
\emph{confounding by indication}, a well known issue in observational healthcare data.

Most realistic causal models contain unmeasured confounding variables that introduce {non-causal} dependencies between treatments and outcomes, and {thus} complicate {establishing valid causal effects}.  For instance, we would expect that in many observational studies the assignment of treatments depends not only on the set of observed baseline characteristics $\vec{C}$, but also unmeasured characteristics $\vec{U}$.  The resulting model is shown {graphically} in Fig.~\ref{fig:backdoor} (b).  Without additional assumptions, the ACE {parameter} $\beta$ is not identified from the observed marginal distribution $p(Y,A,\vec{C})$ in this model \citep{shpitser06id}.

A recent line of work  \citep{miao2018confounding,shi2020multiply,tchetgen2020introduction} has developed a new framework of \emph{proximal causal inference} that allows the use of proxy variables to obtain
identification even in the presence of unobserved confounders.

\subsection{An Example Of Proximal Causal Inference}
\label{subsec:proximal-classic}

To illustrate how proximal causal inference works, we consider a classical observational study with a point exposure $A$, an outcome $Y$, and a set of baseline covariates, some of which are observed ($\vec{C}$), and others are unobserved ($\vec{U}$).  In this study, exposures are assigned based on values of $\vec{C}$ {and} $\vec{U}$.  Thus, the causal relationship of $A$ and $Y$, as quantified by the average causal effect $\beta \equiv \E[Y(a) - Y(a')]$, is obscured by 
non-causal associations induced by {the} covariates.   A causal diagram representing this study is shown in Fig.~\ref{fig:backdoor} (b).

Since some of these covariates are not observed, covariate adjustment will not suffice to identify $\beta$.  Indeed, without further assumptions $\beta$ is not identified from data obtained from such a study.  Proximal causal inference proceeds by assuming that while the vector $\vec{U}$ itself is not observed, there exist observed proxies of $\vec{U}$.  These proxies are subdivided into the control
set which we will call $\vec{Z}$, and the 
{outcome-inducing} 
set, which we will call $\vec{W}$.  Examples of such proxies would include medical tests (ordered both before and after the exposure is assigned) aiming to provide incomplete information about the unobserved course of disease progression in the patients' body.  An alternative name for a control proxy in the literature is a \emph{treatment confounding proxy} or \emph{treatment-inducing proxy}, although the latter name is misleading since, as we will later see, such proxies may either cause, or be caused by treatment variables.

Both proxies may depend on covariates $\vec{C}$, the control proxy $\vec{Z}$ may be associated with the exposure $A$ (in {Fig.~\ref{fig:backdoor} (c)} 
$\vec{Z}$ is caused by $A$, but $\vec{Z}$ may also cause $A$, {as in Fig.~\ref{fig:backdoor} (d)}), while the 
{outcome-inducing} proxy $\vec{W}$ may influence the outcome.
To enable identification, the control and 
{outcome-inducing} proxies must satisfy a number of independence assumptions.  In our {running} example {shown in Fig.~\ref{fig:backdoor} (c)}, since the control proxies $\vec{Z}$ are influenced by the exposure $A$, these assumptions are stated on their counterfactual versions $\vec{Z}(a)$.  However, the assumptions and subsequent derivations carry over without change to Fig.~\ref{fig:backdoor} (d), if $Z(a)$ is replaced by $Z$, as $Z$ is a pre-treatment variable in that model.
{
In fact, if the control proxy $Z$ is a pre-treatment variable, then a stronger identification result is possible, namely that $p(Y(a), \vec{Z})$ is identified, rather than just $p(Y(a))$.  We discuss this further in \S \ref{sec:new}.
}
\begin{assumption}[proxy independence]
\label{a:proxy-id}
{\small
\begin{align}
\label{eqn:assumption-z-i}
\vec{Z}(a) \ci Y(a) &\mid 
{A},\vec{U},\vec{C} \text{ for all }a \in {\mathfrak X}_A\\
\label{eqn:assumption-w}
\vec{W} \ci A,\vec{Z}(a) &\mid \vec{U},\vec{C} \text{ for all }a \in {\mathfrak X}_A
\end{align}
}
\end{assumption}

In addition, the model
{makes assumptions that ensure identification by covariate adjustment had $\vec{U}$ been observed, in addition to $\vec{C}$.}
\begin{assumption}[conditional ignorability]
\label{a:cond-ignore}
{\small
\begin{align}
\label{eqn:ignore}
Y(a) \ci A \mid 
\vec{U},\vec{C} \text{ for all }a \in {\mathfrak X}_A 
\end{align}
}
\end{assumption}
\begin{assumption}[positivity]
\label{a:positivity}
{\small
\begin{align}
\label{eqn:positivity}
p(a \mid \vec{u},\vec{c}) > 0 \text{ for all }a \in{\mathfrak X}_{A},\vec{u}\in{\mathfrak X}_{\vec{U}},\vec{c}\in{\mathfrak X}_{\vec{C}}
\end{align}
}
\end{assumption}

In the interest of conciseness, we will not explicitly mention the positivity assumption in subsequent derivations, and will instead implicitly assume all conditioning events that arise have support.
Similarly, we will implicitly universally quantify independences statements involving counterfactual random variables by all possible intervention values.
{Two} causal diagram{s} that display 
relationships of proxy variables to other variables consistent with the above assumptions 
{are} shown in Fig.~\ref{fig:backdoor} (c) { and }
Fig.~\ref{fig:backdoor} (d).

The link between the unobserved confounder, and the potential outcome of interest is established by means of 
an \emph{outcome bridge function} 
defined as a solution to {an} 
integral equation (specifically a Fredholm equation of the first kind).
\begin{assumption}[outcome bridge function]
\label{a:bridge}
\begin{itemize}
\item[]
\end{itemize}
There exists a bridge function $b(Y,\vec{W},A,\vec{C})$ such that the following equality holds:
{\small
\begin{align}
p(y \mid a,\vec{z},\vec{c}) = \sum_{\vec{w}} b(y,\vec{w},a,\vec{c}) p(\vec{w} \mid a,\vec{c},\vec{z}),
\label{eqn:fredholm}
\end{align}
}
where $\sum_{\vec{w}}$ may denote integration for continuous state spaces.
\end{assumption}

If we are interested in only $\beta$, it is possible to assume an outcome bridge function $\tilde{b}(\vec{W},A,\vec{C})$ that is not a function of $\vec{Y}$ and solves the following equation instead:
{\small
\begin{align*}
\E[Y \mid a,\vec{z},\vec{c}] = \sum_{\vec{w}} \tilde{b}(\vec{w},a,\vec{c}) p(\vec{w} \mid a,\vec{c},\vec{z}).
\end{align*}
}
Nevertheless, we will present integral equations for the density, as in (\ref{eqn:fredholm}) to aid subsequent developments.
Techniques for solving the integral equations of the form in (\ref{eqn:fredholm}) have been derived in functional analysis.  Sufficient conditions for existence of a solution are discussed in detail in the context of proximal causal identification in \citep{miao2018confounding,shi2020multiply,tchetgen2020introduction}.  Necessary and sufficient conditions for the existence of solutions to such integral equations are well understood, see \citep{kress99linear}.
Note that it is not necessary to require that the bridge function be unique.


Finally, we assume {a} 
\emph{completeness condition}.
\begin{assumption}[completeness]
\label{a:complete}
\begin{itemize}
\item[]
\end{itemize}
For any random function $v(\vec{U})$ in $L^2$,
{\small
\begin{align}
\E[v(\vec{U}) \mid \vec{z},a,\vec{c}] = 0 \text{ for all }\vec{z},a \text{ and }\vec{c} \text{ if and only if } v(\vec{U}) = 0, \text{ almost surely}.
\label{eqn:completeness}
\end{align}
}
\end{assumption}
Completeness conditions such as the one in above display have been used extensively in the causal inference literature, particularly in the nonparametric instrumental variable setting \citep{darolles11nonparametric,horowitz11applied,newey03instrumental}.
Completeness can accommodate both categorical and continuous confounders, and corresponds to the requirement that any variation in $\vec{Z}$ can be induced by variation in $\vec{U}$, a formalization of $\vec{U}$-relevance, thus ruling out conditional independence of $\vec{U}$ and $\vec{Z}$ given $A$ and $\vec{C}$. The condition is most intuitive in the case of categorical $\vec{U}$, $\vec{Z}$ and $\vec{W}$, with number of categories $d_{\vec{u}}$, $d_{\vec{z}}$ and $d_{\vec{w}}$ respectively. In this case, completeness requires that
	$min\left( d_{\vec{z}},d_{\vec{w}}\right)  \geq d_{\vec{u}}$.
In other words, $\vec{Z}$ and $\vec{W}$ must each have at least as many categories as $\vec{U}$.  This motivates measuring a rich set of proxies as a potential strategy for mitigating unmeasured
confounding via the proximal approach we now describe. Additional discussion regarding completeness condition can be found in \citep{miao2018confounding,shi2020multiply,tchetgen2020introduction}.

{
In the subsequent discussion, and throughout the remainder of the paper, we will make use of the \emph{{semi-}graphoid axioms} of conditional independence, 
which we reproduce here for completeness \citep{dawid79cond,pearl86graphoids}:
{\small
\begin{align*}
(\vec{A} \ci \vec{B} \mid \vec{C}) \Rightarrow (\vec{B} \ci \vec{A} \mid \vec{C}) & \hspace{0.5cm} \text{(symmetry)}\\
(\vec{A} \ci \vec{B} \cup \vec{D} \mid \vec{C}) \Rightarrow
(\vec{A} \ci \vec{B} \mid \vec{C}) & \hspace{0.5cm} \text{(decomposition)}\\
(\vec{A} \ci \vec{B} \cup \vec{D} \mid \vec{C}) \Rightarrow
(\vec{A} \ci \vec{B} \mid \vec{C} \cup \vec{D}) & \hspace{0.5cm} \text{(weak union)}\\
(\vec{A} \ci \vec{B} \mid \vec{C} \cup \vec{D})
\text{ and } (\vec{A} \ci \vec{D} \mid \vec{C}) \Rightarrow (\vec{A} \ci \vec{B} \cup \vec{D} \mid \vec{C}) & \hspace{0.5cm} \text{(contraction)}
\end{align*}
}
where $\vec{A},\vec{B},\vec{C},\vec{D}$ are sets of random variables.
}

If (\ref{eqn:assumption-z-i}) and (\ref{eqn:assumption-w}) are assumed to hold, $\vec{U}$-relevance implied by the completeness condition (\ref{eqn:completeness}) is a necessary condition for the existence of solutions to the integral equations for the bridge function.
This is because if $(\vec{W} \ci \vec{U} \mid \vec{C})$ {holds}, (\ref{eqn:assumption-w}) and the contraction graphoid axiom implies that
$(\vec{W} \ci A,\vec{Z}(a),\vec{U} \mid \vec{C})$ holds, {whereas if $(\vec{Z} \ci \vec{U} \mid A,\vec{C})$ holds, (\ref{eqn:assumption-z-i}) and the contraction graphoid axiom implies that
$(\vec{Z} \ci Y,\vec{U} \mid A{ =a},\vec{C})$ holds.}
Either of these independences further imply the above equations do not admit solutions.  Thus, all identification results that rely on existence of solutions of Fredholm equations such as (\ref{eqn:fredholm}) are examples of \emph{generic} identification, and will not hold in all causal models corresponding to the graphs in Fig.~\ref{fig:backdoor} (c) or (d), but only in those models where appropriate dependences hold between unobserved confounders and proxy variables.


$\vec{U}$-relevance may be viewed as a form of \emph{faithfulness} \citep{spirtes01causation}, but only with respect to edges between $\vec{U}$ and proxies $\vec{Z},\vec{W}$ implying the above dependences.
Note that, similar to the proximal causal models we consider here, the well-established instrumental variable model is likewise a framework for generic identification, and similarly imposes a partial faithfulness requirement in the sense that the instrument and the treatment variables must be associated for identification to hold.

Given the above assumptions, the counterfactual mean $\E[Y(a)]$ (and therefore $\beta$) is identified.  Specifically, we have the following result, with the proof included for illustrative purposes.
\begin{theorem}[proximal g-formula, \citep{miao2018identifying}]
Under assumptions \ref{a:proxy-id}, \ref{a:cond-ignore}, \ref{a:positivity}, \ref{a:bridge}, and \ref{a:complete},
{\small
\begin{align*}
p(Y(a)) = \sum_{\vec{c},\vec{w}} b(y,\vec{w},a,\vec{c}) p(\vec{w},\vec{c}).
\end{align*}
}
\label{thm:proximal-backdoor}
\end{theorem}
\vspace{-0.7cm}
\begin{proof}
We first derive the following identities:
{\small
\begin{align}
(\ref{eqn:assumption-z-i})
& \underbrace{\Rightarrow}_{\text{consistency}}
\label{eqn:assumption-z}
Y \ci \vec{Z} \mid \vec{U},\vec{C},A=a.\\
\label{eqn:assumption-w-2}
(\ref{eqn:assumption-w})
&\underbrace{\Rightarrow}_{\text{weak union}}
\vec{W} \ci \vec{Z}(a) \mid A=a,\vec{U},\vec{C} \underbrace{\Rightarrow}_{\text{consistency}} \vec{W} \ci \vec{Z} \mid A=a,\vec{U},\vec{C}\\
\label{eqn:assumption-w-3}
(\ref{eqn:assumption-w})
&\underbrace{\Rightarrow}_{\text{decomposition}}
\vec{W} \ci A \mid \vec{U},\vec{C}
\end{align}
}
We proceed as follows:
{\small
\begin{align}
\notag
p(Y \mid a,\vec{z},\vec{c})
&= \sum_{\vec{w}} b(y,\vec{w},a,\vec{c}) p(\vec{w} \mid a,\vec{c},\vec{z}) \Rightarrow (\text{by }(\ref{eqn:assumption-z}),(\ref{eqn:assumption-w-2}))\\
\notag
\sum_{\vec{u}} p(Y \mid a,\vec{u},\vec{c}) p(\vec{u} \mid a,\vec{z},\vec{c})
&= \sum_{\vec{w}} b(y,\vec{w},a,\vec{c}) \sum_{\vec{u}} p(\vec{w} \mid a,\vec{c},\vec{u}) p(\vec{u} \mid a,\vec{z},\vec{c}) \Rightarrow (\text{by }(\ref{eqn:completeness}))\\
\notag
p(Y \mid a,\vec{u},\vec{c})
&= \sum_{\vec{w}} b(y,\vec{w},a,\vec{c}) p(\vec{w} \mid a,\vec{c},\vec{u}) \Rightarrow (\text{by }(\ref{eqn:assumption-w-3}))\\
\notag
\sum_{\vec{u},\vec{c}} p(Y \mid a,\vec{u},\vec{c}) p(\vec{u},\vec{c})
&= \sum_{\vec{u},\vec{c}} \sum_{\vec{w}} b(y,\vec{w},a,\vec{c}) p(\vec{w} \mid \vec{c},\vec{u}) p(\vec{u},\vec{c}) \Rightarrow (\text{by }(\ref{eqn:ignore}))\\
p(Y(a))
&= \sum_{\vec{c},\vec{w}} b(y,\vec{w},a,\vec{c}) p(\vec{w},\vec{c}).
\label{eqn:p-g}
\end{align}
}
{This establishes the result.}
\end{proof}

The functional in (\ref{eqn:p-g}) was called the \emph{proximal g-formula} in \citep{tchetgen2020introduction}.  Estimation methods for this functional have been discussed in \citep{tchetgen2020introduction,semiparametric2020cui}.

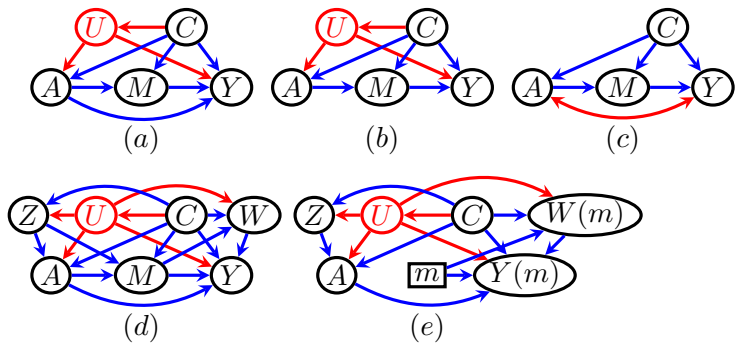
\begin{figure}[!h]
\centering
\begin{tikzpicture}[>=stealth, node distance=1.2cm]
    \tikzstyle{format} = [draw, very thick, ellipse,
                          minimum size=0.2cm, inner sep=1pt]
    \tikzstyle{format1} = [draw, very thick, ellipse,
                          minimum size=0.2cm, inner sep=1pt]
    \tikzstyle{format0} = [draw, very thick, ellipse,
                          minimum size=0.2cm, inner sep=0pt]
    \tikzstyle{unobs} = [draw, very thick, red, ellipse,
                         minimum size=0.2cm, inner sep=2pt]
    \tikzstyle{square} = [draw, very thick, rectangle,
                         minimum size=0.2cm, inner sep=2pt]

    \begin{scope}[xshift=3.2cm]
        \path[->, very thick]
            node[format] (A) {$A$}
            node[format1, right of=A] (M) {$M$}
            node[format, right of=M] (Y) {$Y$}
            node[format, red, above of=M, xshift=-0.6cm, yshift=-0.4cm] (U) {${U}$}
            node[format, above of=M, xshift=0.6cm, yshift=-0.4cm] (C) {${C}$}

		(C) edge[red] (U)

		(U) edge[red] (A)
		(U) edge[red] (Y)
		(C) edge[blue] (A)
		(C) edge[blue] (M)
		(C) edge[blue] (Y)

		(A) edge[blue] (M)
		(M) edge[blue] (Y)

            node[below of=M, yshift=0.5cm] (l) {$(b)$}
        ;
    \end{scope}

    \begin{scope}[xshift=6.4cm]
        \path[->, very thick]
            node[format] (A) {$A$}
            node[format1, right of=A] (M) {$M$}
            node[format, right of=M] (Y) {$Y$}
            node[format, above of=M, xshift=0.6cm, yshift=-0.4cm] (C) {${C}$}


		(A) edge[<->, red, bend right=28] (Y)
		(C) edge[blue] (A)
		(C) edge[blue] (M)
		(C) edge[blue] (Y)

		(A) edge[blue] (M)
		(M) edge[blue] (Y)

            node[below of=M, yshift=0.5cm] (l) {$(c)$}
        ;
    \end{scope}
    
    \begin{scope}
        \path[->, very thick]
            node[format] (A) {$A$}
            node[format1, right of=A] (M) {$M$}
            node[format, right of=M] (Y) {$Y$}
            node[format, red, above of=M, xshift=-0.6cm, yshift=-0.4cm] (U) {${U}$}
            node[format, above of=M, xshift=0.6cm, yshift=-0.4cm] (C) {${C}$}

		(C) edge[red] (U)

		(U) edge[red] (A)
		(U) edge[red] (Y)
		(C) edge[blue] (A)
		(C) edge[blue] (M)
		(C) edge[blue] (Y)

		(A) edge[blue] (M)
		(M) edge[blue] (Y)

		(A) edge[blue, bend right] (Y)

            node[below of=M, yshift=0.5cm] (l) {$(a)$}
        ;
    \end{scope}
    
    \begin{scope}[yshift=-2.5cm, xshift=0cm]
        \path[->, very thick]
            node[format] (A) {$A$}
            node[format1, right of=A] (M) {$M$}
            node[format, right of=M] (Y) {$Y$}
            node[format, red, above of=M, xshift=-0.6cm, yshift=-0.4cm] (U) {${U}$}
            node[format, above of=M, xshift=0.6cm, yshift=-0.4cm] (C) {${C}$}

		node[format, xshift=0.3cm, left of=U] (Z) {$Z$}
		node[format1, xshift=-0.3cm, right of=C] (W) {$W$}

		(C) edge[red] (U)

		(U) edge[red] (A)
		(U) edge[red] (Y)
		(C) edge[blue] (A)
		(C) edge[blue] (M)
		(C) edge[blue] (Y)

		(A) edge[blue] (M)
		(M) edge[blue] (Y)
		
		(Z) edge[blue] (A)
		(Z) edge[blue] (M)

		(M) edge[blue] (W)
		(W) edge[blue] (Y)
		
		(U) edge[red] (Z)
		(U) edge[red, bend left] (W)
		
		(C) edge[blue] (W)
		(C) edge[blue, bend right] (Z)
		
		(A) edge[blue, bend right=28] (Y)

            node[below of=M, yshift=0.5cm] (l) {$(d)$}
        ;
    \end{scope}
  
      \begin{scope}[yshift=-2.5cm, xshift=3.8cm]
        \path[->, very thick]
            node[format] (A) {$A$}
            node[square, right of=A] (M) {$m$}
            node[format0, right of=M, xshift=0.1cm] (Y) {$Y(m)$}
            node[format, red, above of=M, xshift=-0.6cm, yshift=-0.4cm] (U) {${U}$}
            node[format, above of=M, xshift=0.6cm, yshift=-0.4cm] (C) {${C}$}

		node[format, xshift=0.3cm, left of=U] (Z) {$Z$}
		node[format0, xshift=0.3cm, right of=C] (W) {$W(m)$}

		(C) edge[red] (U)

		(U) edge[red] (A)
		(U) edge[red] (Y)
		(C) edge[blue] (A)
		(C) edge[blue] (Y)

		(M) edge[blue] (Y)
		
		(Z) edge[blue] (A)

		(M) edge[blue] (W)
		(W) edge[blue] (Y)
		
		(U) edge[red] (Z)
		(U) edge[red, bend left] (W)
		
		(C) edge[blue] (W)
		(C) edge[blue, bend right] (Z)
		
		(A) edge[blue, bend right=25] (Y)

            node[below of=M, yshift=0.5cm] (l) {$(e)$}
        ;
    \end{scope}
      
\end{tikzpicture}
\caption{
(a) A causal diagram representing a model where nonparametric point identification fails due to the presence of the unobserved common cause of $A$ and $Y$ and the existence of the direct effect of $A$ on $Y$.
(b) A causal diagram representing a model where nonparametric point identification of the population ACE is possible via the front-door criterion of Pearl.
(c) A latent projection ADMG of the DAG in (b).
(d) A causal diagram representing a model where point identification of the population ACE is impossible nonparametrically, but possible with proximal causal inference methods.
(e) A causal diagram representing a world where the mediator variable $M$ is intervened on to value $m$, resulting in a hypothetical world where proximal causal inference methods may be used to
identify $\E[Y(a,m)]$.
}
\label{fig:front-door}
\end{figure}

\section{The Proximal Front-Door Criterion}
\label{sec:proximal-2}

The basic example of proximal causal inference presented in the previous section relies on assumptions that will not always be satisfied in practical applications.
For instance, consider the model shown in Fig.~\ref{fig:front-door} (a), which resembles the setting with unobserved confounding shown in Fig.~\ref{fig:backdoor} (b), but with an additional complication that the causal relationship of the treatment $A$ and outcome $Y$ is partially mediated by an observed variable $M$, which may potentially depend on observed covariates $\vec{C}$.

We now illustrate how additional assumptions may render the ACE parameter $\beta = \E[Y(a) - Y(a')]$ identified in the submodel of Fig.~\ref{fig:front-door} (a) shown in Fig.~\ref{fig:front-door} (b).
The first assumption we will use is that $\vec{M}$ is a \emph{strong mediator}, or that interventions on $A$ have no effect on $Y$, provided $\vec{M}$ is also intervened on, or:
{\small
\begin{align}
\label{eqn:strong-med}
Y(a,\vec{m}) = Y(\vec{m})\text{ for all }a,\vec{m}.
\end{align}
}
In addition, we will use the following assumptions:
{\small
\begin{align}
\label{eqn:ignore-y-m}
Y(a,\vec{m}) \ci \vec{M}(a) \mid \vec{C}\\
\label{eqn:ignore-y-m-2}
Y(\vec{m}) \ci M \mid \vec{C},A\\
\label{eqn:ignore-m-a}
\vec{M}(a) \ci A \mid \vec{C}.
\end{align}
}
The submodel of Fig.~\ref{fig:front-door} (a) characterized by the above assumptions yields nonparametric identification of $p(Y(a))$ via Pearl's \emph{front-door formula}:
{\small
\begin{align}
p(Y(a)) = \sum_{\vec{c},\vec{m}} \left( \sum_{\tilde{a}} p(Y \mid \tilde{a},\vec{m},\vec{c}) p({\tilde{a}} \mid \vec{c}) \right) p(\vec{m} \mid a,\vec{c}) p(\vec{c}),
\label{eqn:front-door}
\end{align}
}
via the following derivation: 
{\small
\begin{align*}
p(Y(a)) = 
p(Y(a,\vec{M}(a)))
&= \sum_{\vec{m},\vec{c}} p(Y(a,\vec{m}) \mid \vec{M}(a) = \vec{m},\vec{c}) p(\vec{M}(a)=\vec{m} \mid \vec{c}) p(\vec{c}) \Rightarrow (\text{by }(\ref{eqn:ignore-y-m}))\\
&= \sum_{\vec{m},\vec{c}} p(Y(a,\vec{m}) \mid \vec{c}) p(\vec{M}(a)=\vec{m} \mid \vec{c}) p(\vec{c}) \Rightarrow(\text{by }(\ref{eqn:strong-med}))\\
&= \sum_{\vec{m},\vec{c}} p(Y(\vec{m}) \mid \vec{c}) p(\vec{M}(a)=\vec{m} \mid \vec{c}) p(\vec{c}) \Rightarrow(\text{by }(\ref{eqn:ignore-y-m-2}),(\ref{eqn:ignore-m-a}))\\
&= \sum_{\vec{m},\vec{c}} \left( \sum_{{\tilde{a}}} p(Y \mid \tilde{a},\vec{m},\vec{c}) p(\tilde{a} \mid \vec{c}) \right) p(\vec{m} \mid a, \vec{c}) p(\vec{c})
\end{align*}
}

The strong mediator assumption (\ref{eqn:strong-med}) is often not realistic in applications, as it implies the analyst is able to find a set of observed variables $\vec{M}$ that completely mediate the effect of $A$ on $Y$.  In settings where finding such a set is unrealistic, the above identification strategy is dubious.  An alternative approach to obtaining identification is to use proximal causal inference.
This would entail positing, as in the
model in Fig.~\ref{fig:backdoor} (c) discussed above, a set $\vec{Z}$ of {control proxies} of $\vec{U}$, and a set $\vec{W}$ of 
{outcome-inducing} proxies of $\vec{U}$, resulting in a causal diagram shown in Fig.~\ref{fig:front-door} (d).
Note that unlike 
the model shown in Fig.~\ref{fig:backdoor} (c), the presence of the mediator $M$ prevents assumptions (\ref{eqn:assumption-w}) and {thus restriction} 
(\ref{eqn:assumption-w-2})
from holding {in Fig.~\ref{fig:front-door} (d)}.
Nevertheless, it is possible to obtain identification by noting that if we view $\vec{M}$ as a treatment, it only has observed direct causes, meaning that the effect of $\vec{M}$ on all other {observed} variables may be obtained by the g-formula:
{\small
\begin{align}
\label{eqn:g-m}
p^{(\vec{m})}(Y(\vec{m}),\vec{W}(\vec{m}), \vec{Z},A,\vec{C})
= p(Y,\vec{W}
\mid m, A, \vec{Z}{,\vec{C}}) p(A,\vec{Z}{,\vec{C}}).
\end{align}
}Formally, this is justified by appealing to a version of the conditional ignorability assumption (\ref{eqn:ignore-m}), below, which holds in this model.

The identifying functional for $p^{(\vec{m})}$ is a 
\emph{Markov kernel} \citep{lauritzen96graphical}, an object that behaves like a conditional distribution but is not necessarily obtained by a conditioning operation.  Independence restrictions in {the Markov kernel corresponding to} $p^{(\vec{m})}$
may be displayed by the d-separation criterion in a \emph{conditional DAG} shown in Fig.~\ref{fig:front-door} (e), {obtained from Fig.~\ref{fig:front-door} (d) by removing} all arrows into $M$ 
and viewing $M$ {as} 
``fixed'' (which we denote by a rectangle rather than an ellipse) meaning that 
$M$ no longer corresponds to a random variable.

It turns out that we can restate assumptions needed for identification that held in the observed data distribution $p$ in \S \ref{subsec:proximal-classic} for the kernel $p^{(\vec{m})}$, as follows.

\begin{assumption}[kernel proxy independence]
\label{a:k-proxy-id}
{\small
\begin{align}
\label{eqn:assumption-z-q-i}
\vec{Z}(
\vec{m}) \ci Y(a,\vec{m}) \mid 
{A(\vec{m}) , }
\vec{U}(\vec{m}) ,\vec{C}(\vec{m})\\
\label{eqn:assumption-w-q}
\vec{W}(\vec{m})  \ci A(\vec{m}) , \vec{Z}(
\vec{m})  \mid \vec{U}(\vec{m}) ,\vec{C}(\vec{m})
\end{align}
}
\end{assumption}
In the independence statements above, we label all variables as potential outcomes after an intervention on $\vec{m}$ is performed to make clear that they all hold in $p^{(\vec{m})}$, regardless of whether the variable actually depends on the intervened on values $\vec{m}$.

However, since $A,\vec{U},\vec{C}$ and $\vec{Z}$ precede $M$ in any causal ordering consistent with Fig.~\ref{fig:front-door} (d), these variables do not actually depend on the intervened on values of $m$.  This is represented by the following individual level restrictions.
\begin{assumption}[exclusion restrictions]
\label{a:exclusion-1}
\begin{itemize}
\item[]
\end{itemize}
{\small
\begin{align}
A(\vec{m}) = A, \vec{U}(\vec{m}) = \vec{U}, \vec{C}(\vec{m}) = \vec{C},\text{ and }\vec{Z}(
\vec{m}) = {\vec{Z}}
.
\label{eqn:exclusion-1}
\end{align}
}
\end{assumption}
The generalizations of assumption \ref{a:cond-ignore} that holds in the model in Fig.~\ref{fig:front-door} (d) is shown below.  This assumption holds for $Y(a,m)$ and $A$ only conditional on $\vec{U}$ 
{and} $\vec{C}
$, and in a hypothetical situation where an intervention on $M$ took place.  However, additional ignorability assumptions relating $M$ and $W,Y$ and $A$ are also needed.  These assumptions must hold without conditioning on $\vec{U}$ (this is illustrated graphically by $M$ not having $\vec{U}$ as a parent in Fig.~\ref{fig:front-door} (d)).  In other words, to obtain identification we allow the relaxation of restriction
(\ref{eqn:assumption-w})
while requiring additional assumptions on the mediator variable $M$.
\begin{assumption}[generalized conditional ignorability]
\label{a:k-cond-ignore}
{\small
\begin{align}
\label{eqn:ignore-y-m-a}
Y(a,\vec{m})
\ci 
A \mid \vec{U},\vec{C}
\\
\label{eqn:ignore-m}
{Y(\vec{m}),}\vec{W}(\vec{m})
\ci \vec{M} \mid { A,}\vec{C},\vec{Z}\\
\label{eqn:ignore-a-no-u}
\vec{M}(a) \ci Y(a,\vec{m}), A \mid C, \vec{Z}
.
\end{align}
}
\end{assumption}
The independence restrictions above can be read off either from
Fig.~\ref{fig:front-door} (e), or a generalization {based on Single World Intervention Graphs (SWIGs)} described in \citep{thomas13swig}.

The following two conditions generalize assumptions \ref{a:bridge} and \ref{a:complete}, and hold in the distribution $p^{(m)}$.
\begin{assumption}[kernel outcome bridge function]
\label{a:k-bridge}
\begin{itemize}
\item[]
\end{itemize}
{
There exists a bridge function $b^{(\vec{m})}(Y,\vec{W},A,\vec{C},\vec{m})$ such that the following equality holds:
}
{\small
\begin{align}
p^{(\vec{m})}(Y(\vec{m}) \mid a, \vec{z},\vec{c}) 
&= \sum_{\vec{w}} b^{(\vec{m})}(Y, \vec{w},a,\vec{c},\vec{m}) p^{(\vec{m})}(\vec{W}(\vec{m}) \mid a,\vec{c},\vec{z})
\label{eqn:fredholm-q}
\end{align}
}
\end{assumption}

\begin{assumption}[kernel completeness]
\label{a:k-complete}
\begin{itemize}
\item[]
\end{itemize}
For any random function $v(\vec{U})$ in $L^2$,
{\small
\begin{align}
\E_{p^{(\vec{m})}}[v(\vec{U}) \mid \vec{z},a,\vec{c}
] = 0 \text{ for all }\vec{z},a,\vec{m},\text{ and }\vec{c}\text{ if and only if }v(\vec{U}) = 0, \text{ almost surely}.
\label{eqn:completeness-q-m}
\end{align}
}
\end{assumption}
Note that the conditional expectation in (\ref{eqn:completeness-q-m}) is a function of $\vec{m}$, in addition to $\vec{z},a,\vec{c}$, since the expectation is with respect to an interventional distribution that depends on $\vec{m}$.


{Putting these assumptions together}, we have the following identification result.
{
\begin{theorem}[proximal front-door criterion]
Under assumptions \ref{a:k-proxy-id}, \ref{a:exclusion-1}, \ref{a:k-cond-ignore}, \ref{a:k-bridge}, \ref{a:k-complete}, and positivity,
{\small
\begin{align}
\label{eqn:proximal-frontdoor}
p(Y(a))
&=
\sum_{\vec{m},\vec{c},\vec{z}}
\left(
\sum_{\vec{w}} b^{(\vec{m})}(y,\vec{w},a,\vec{c},\vec{m})
\sum_{\tilde{a}}
p(\vec{w} \mid \vec{m},\tilde{a},\vec{c},\vec{z}) p(\tilde{a} \mid \vec{c},\vec{z})
\right)
p(\vec{m} \mid a, \vec{c},\vec{z}) p(\vec{c},\vec{z}).
\end{align}
}
\label{thm:proximal-frontdoor}
\end{theorem}
}
\begin{proof}
We first obtain the following identities:
{
\small
\begin{align}
(\ref{eqn:assumption-z-q-i})
+
(\ref{eqn:exclusion-1})
&\underbrace{\Rightarrow}_{\text{consistency}}
\label{eqn:assumption-z-q}
Y(\vec{m}) \ci \vec{Z} \mid \vec{U},\vec{C},A=a\\
{
(\ref{eqn:ignore-y-m-a})
+
(\ref{eqn:exclusion-1})
+
(\ref{eqn:assumption-z-q-i})
}
&
{
\underbrace{\Rightarrow}_{\text{contraction}}
\label{eqn:ignore-y-m-a-2}
\vec{Z},A \ci Y(a,\vec{m}) \mid \vec{U} ,\vec{C}
\underbrace{\Rightarrow}_{\text{weak union}}
A \ci Y(a,\vec{m}) \mid \vec{U} ,\vec{C},\vec{Z}
}
\\
\label{eqn:assumption-w-q-2}
(\ref{eqn:assumption-w-q})
+
(\ref{eqn:exclusion-1})
&
\underbrace{\Rightarrow}_{\text{weak union}}
\vec{W}(\vec{m}) \ci \vec{Z}
\mid A
, \vec{U},\vec{C}
\\
\label{eqn:assumption-w-q-3}
(\ref{eqn:assumption-w-q})
+
(\ref{eqn:exclusion-1})
&
\underbrace{\Rightarrow}_{\text{weak union}}
\vec{W}(\vec{m}) \ci A \mid \vec{Z}
, \vec{U},\vec{C}.
\end{align}
}

We are now ready for the following derivation, where we denote $b^{(m)}(y,\vec{w},a,\vec{c},\vec{m})$ by $b^{(m)}$, and $p^{(m)}(Y(\vec{m}) \ldots)$ as $p(Y(\vec{m}) \ldots)$ for conciseness:
{
\small
\begin{align}
\notag
p(Y(\vec{m}) \mid a, \vec{z},\vec{c})\
&= \sum_{\vec{w}} b^{(\vec{m})}
p(\vec{W}(\vec{m}){=\vec{w}} \mid a,\vec{z},\vec{c})
\Rightarrow (\text{by }(\ref{eqn:assumption-z-q}),(\ref{eqn:assumption-w-q-2}))\\
\notag
\sum_{\vec{u}} p(Y(\vec{m}) \mid a,\vec{u},\vec{c}) p(\vec{U}=\vec{u} \mid a,\vec{z},\vec{c})
&= \sum_{\vec{w}} b^{(\vec{m})}
\sum_{\vec{u}} p(\vec{W}(\vec{m})=\vec{w} \mid a,\vec{u},\vec{c}) p(\vec{U}=\vec{u} \mid a,\vec{z},\vec{c}) \Rightarrow\\
\notag
& \hspace{7.1cm} (\text{by }(\ref{eqn:completeness-q-m}))\\
\notag
p(Y(\vec{m}) \mid a,\vec{u},\vec{c})
&=\sum_{\vec{w}} b^{(\vec{m})}
p(\vec{W}(\vec{m})=\vec{w} \mid a,\vec{u},\vec{c}) \Rightarrow (\text{by }(\ref{eqn:assumption-z-q}),(\ref{eqn:assumption-w-q-2}),(\ref{eqn:assumption-w-q-3}))\\
\notag
\sum_{\vec{u}} p(Y(\vec{m}) \mid a,\vec{u},\vec{c}, \vec{z}) p(\vec{U}=\vec{u} \mid \vec{c},\vec{z})
&= \sum_{\vec{u}} \sum_{\vec{w}} b^{(\vec{m})}
p(\vec{W}(\vec{m})=\vec{w} \mid \vec{u},\vec{c},\vec{z}) p(\vec{U}=\vec{u} \mid \vec{c},\vec{z}) \Rightarrow\\
\notag
&
\hspace{0.5cm}
(\text{by }
(\ref{eqn:ignore-y-m-a-2})
\text{ and consistency: 
$Y(m,a)=Y(m)$ if $A=a$})\\
\label{eqn:final-front}
p(Y(a,\vec{m}) \mid \vec{c},\vec{z})
&= \sum_{\vec{w}} b^{(\vec{m})}
p(\vec{W}(\vec{m})=\vec{w} \mid \vec{c},\vec{z}).
\end{align}
}
{
To identify $p(Y(a))$, we note that
{\small
\begin{align}
\notag
p(Y(a))
&= \sum_{\vec{c},\vec{z}} p(Y(a) \mid \vec{c},\vec{z}) p(\vec{c},\vec{z})
= \sum_{\vec{c},\vec{z}} p(Y(a,\vec{M}(a)) \mid \vec{c},\vec{z}) p(\vec{c},\vec{z})\\
\notag
&= \sum_{\vec{m},\vec{c},\vec{z}} p(Y(a,\vec{m}), \vec{M}(a)=\vec{m} \mid \vec{c},\vec{z}) p(\vec{c},\vec{z})\\
\notag
&= \sum_{\vec{m},\vec{c},\vec{z}} p(Y(a,\vec{m}) \mid \vec{c},\vec{z}) p(\vec{M}(a)=\vec{m} \mid \vec{c},\vec{z}) p(\vec{c},\vec{z}) \quad (\text{by (\ref{eqn:ignore-a-no-u})}) \\
\notag
&= \sum_{\vec{m},\vec{c},\vec{z}} \left( \sum_{\vec{w}} b^{(\vec{m})} p(\vec{W}(\vec{m})=\vec{w} \mid \vec{c},\vec{z}) \right) p(\vec{M}(a)=\vec{m} \mid \vec{c},\vec{z}) p(\vec{c},\vec{z})
\quad (\text{by (\ref{eqn:final-front})})\\
&= \sum_{\vec{m},\vec{c},\vec{z}} \left( \sum_{\vec{w}} b^{(\vec{m})} p(\vec{W}(\vec{m})=\vec{w} \mid \vec{c},\vec{z}) \right) p(\vec{m} \mid a,\vec{c},\vec{z}) p(\vec{c},\vec{z})
\quad (\text{by (\ref{eqn:ignore-a-no-u})}).
\label{eqn:front-door-mid}
\end{align}
}
Since 
$p(Y(\vec{m}),\vec{W}(\vec{m}), \vec{Z},A,\vec{C})$ is identified by (\ref{eqn:g-m}) by appealing to (\ref{eqn:ignore-m}),
$p(\vec{W}(\vec{m})=\vec{w} \mid \vec{c},\vec{z})$ is also identified (via conditioning and marginalization) as 
$\sum_{\tilde{a}} p(\vec{w}\mid m, \tilde{a}, \vec{z},\vec{c}) p(\tilde{a} \mid \vec{z},\vec{c})$.
Plugging this expression into (\ref{eqn:front-door-mid}) yields our conclusion:
{\small
\begin{align*}
p(Y(a))
=
\sum_{\vec{m},\vec{c},\vec{z}} \left( \sum_{\vec{w}} b^{(\vec{m})}(y,\vec{w},a,\vec{c},\vec{m}) \sum_{\tilde{a}} p(\vec{w}\mid m, \tilde{a}, \vec{z},\vec{c}) p(\tilde{a} \mid \vec{z},\vec{c}) \right) p(\vec{m} \mid a,\vec{c},\vec{z}) p(\vec{c},\vec{z}).
\end{align*}
}
}

\end{proof}
{
Note that since (\ref{eqn:ignore-m}) only holds conditional on $\vec{Z}$, it was important for the overall derivation to work that the distribution identified by (\ref{eqn:final-front}) conditioned on $\vec{Z}$, which in turn was only possible because the control proxies in $\vec{Z}$ were pre-treatment.
}

We describe an approach to estimating the target parameter identified by (\ref{eqn:proximal-frontdoor}) in \S \ref{sec:sims}.
We compare this method against baselines defined using the proximal g-formula from (\ref{eqn:p-g}) and the front-door from (\ref{eqn:front-door}). By changing our synthetic data distributions, we show that the presence of an $A \to Y$ edge violates assumption (\ref{eqn:strong-med}) of the front-door method, and that a $Z \to M \to W$ path violates assumptions (\ref{eqn:assumption-w}) and
(\ref{eqn:assumption-z}) of the proximal g-formula. These simulations give empirical evidence for the importance of using proximal methods when the assumptions of simpler approaches are violated.

\section{The ID Algorithm}
\label{sec:id}

General nonparametric identification results, of which the front-door formula (\ref{eqn:front-door}) is a special case, have been derived using the machinery of graphical causal models, culminating in the characterization of nonparametric identification via the ID algorithm \citep{tian02on,shpitser06id,huang06do}.  Here we briefly review this background, prior to introducing a general approach to synthesizing proximal causal inference and nonparametric identification.

The statistical model of a directed acyclic graph (DAG) ${\cal G}(\vec{V})$ with a vertex set $\vec{V} \equiv \{ V_1, \ldots, V_k \}$, called a \textit{Bayesian network}, is the set of distributions that Markov factorize with respect to the DAG as $p(\vec{V}) = \prod_{V_i \in \vec{V}} p(V_i \mid \pa_{\cal G}(V_i))$ where $\pa_{\cal G}(V_i)$ are parents of $V_i$ in ${\cal G}$.  {Directed acyclic graphs have also been used to represent \emph{causal models}, which describe cause-effect relationships between variables, as follows.}

Each variable $V_i$ in a causal model {of a DAG} is determined from values of its parents $\pa_{\cal G}(V_i)$ and an exogenous noise variable $\epsilon_i$ via an invariant causal mechanism called a \emph{structural equation} $f_i(\pa_{\cal G}(V_i), \epsilon_i)$.  Causal models allow counterfactual intervention operations, denoted by the $\text{do}(\vec{A})$ operator in
\citep{pearl09causality}.  Such operations replace each structural equation $f_i(\pa_{\cal G}(V_i), \epsilon_i)$ for $V_i \in \vec{A} \subset \vec{V}$ by one that sets $V_i$ to a constant value in $\vec{A}$ corresponding to $V_i$.  The joint distribution of variables in $\vec{Y} \equiv \vec{V} \setminus \vec{A}$ after the intervention $\text{do}(\vec{A})$ was performed is denoted by
$p(\vec{Y} \mid \text{do}(\vec{a}))$, equivalently written as $p(\{ V_i(\vec{a}) : V_i \in \vec{Y} \})$, or $p(\vec{Y}({\vec{a}}))$, where $V_i(\vec{a})$
is a counterfactual random variable or a potential outcome.

A popular causal model called the \emph{structural causal model (SCM)} or the \emph{nonparametric structural equation model with independent errors (NPSEM-IE)} \citep{pearl09causality} assumes, aside from the structural equations for each variable being functions of their parents in the DAG ${\cal G}(\vec{V})$, that 
the joint distribution of all exogenous terms
{admits the following factorization into marginal distributions}: $p(\{ \epsilon_i : V_i \in \vec{V} \}) = \prod_{V_i \in \vec{V}} p(\epsilon_i)$.  The NPSEM-IE implies the DAG factorization of $p(\vec{V})$ with respect to ${\cal G}(\vec{V})$, and a truncated DAG factorization known as the \emph{g-formula}:

{\small 
\begin{align}
p(\vec{Y}(\vec{a})) =
\prod_{V_i \in \vec{Y}} p(V_i \mid \pa_{\cal G}(V_i))
\vert_{ \vec{A} = \vec{a} }
\label{eqn:g}
\end{align}
}
for every $\vec{A} \subseteq \vec{V}$, and $\vec{Y} = \vec{V} \setminus \vec{A}$.
{
Here $p(V_i \mid \pa_{\cal G}(V_i))\vert_{ \vec{A} = \vec{a} }$ means ``evaluate
$p(V_i \mid \pa_{\cal G}(V_i))$ at values $a_i \in \vec{a}$, for every $A_i \in \vec{A}$ that appears in $\pa_{\cal G}(V_i)$.''
}

The g-formula (\ref{eqn:g}) provides an elegant link between observed data and counterfactual distributions in causal models where all relevant variables are observed.
Causal models that arise in practice, however,  
contain hidden variables.  
Representing such models using a DAG ${\cal G}(\vec{V} \cup \vec{H})$ where 
$\vec{V}$ and $\vec{H}$ correspond to observed and hidden variables, respectively,
is not very helpful, since 
applying (\ref{eqn:g}) to ${\cal G}(\vec{V} \cup \vec{H})$ results in an expression that involves unobserved variables $\vec{H}$.
A popular alternative is to represent a class of hidden variable DAGs ${\cal G}_i(\vec{V} \cup \vec{H}_i)$ by a single
\emph{acyclic directed mixed graph} (ADMG) ${\cal G}(\vec{V})$ that contains directed ($\to$) and bidirected ($\leftrightarrow$) edges and no directed cycles
via the \emph{latent projection} operation \citep{verma90equiv}. 

Identification theory of all interventional distributions $p(\vec{Y}(\vec{a}))$ in a hidden variable causal model associated with a DAG ${\cal G}(\vec{V} \cup \vec{H})$
may be expressed on the latent projection ADMG ${\cal G}(\vec{V})$ without loss of generality \citep{richardson17nested}.

Under such a causal model, for any disjoint $\vec{A},\vec{Y} \subseteq \vec{V}$, $p(\vec{Y}(\vec{a}))$ is identified if and only if $p(\vec{Y}^*(\vec{a}))$ is identified, where $\vec{Y}^*$ is the set of ancestors of $\vec{Y}$ in ${\cal G}(\vec{V})$ via directed paths not through $\vec{A}$. The distribution $p(\vec{Y}^*(\vec{a}))$ factorizes with respect to a graph ${\cal G}(\vec{V})_{{\vec Y}^*}$, obtained from ${\cal G}(\vec{V})$ retaining only vertices in $\vec{Y}^*$ and edges between these vertices.  The factorization of $p(\vec{Y}^*(\vec{a}))$ is in terms of a set of interventional distributions associated with bidirected connected sets in ${\cal G}(\vec{V})_{{\vec Y}^*}$.  These sets are called \emph{districts}, with the set of all districts in ${\cal G}(\vec{V})_{{\vec Y}^*}$ denoted by ${\cal D}({\cal G}(\vec{V})_{{\vec Y}^*})$.

In particular, we have:
{\small
	\setlength{\abovedisplayskip}{5pt} \setlength{\belowdisplayskip}{5pt}
	\begin{align}
	p(\vec{Y}(\vec{a})) = \sum_{\vec{Y}^* \setminus \vec{Y}} p(\vec{Y}^*(\vec{a})) =
	\sum_{\vec{Y}^* \setminus \vec{Y}} \prod_{\vec{D} \in {\cal D}({\cal G}{(\vec{V})}_{\vec{Y}^*})}
		p(\vec{D} \mid \text{do}(\vec{s}_{\vec D}))
	\label{eqn:id}
	\end{align}
}
where $\vec{s}_{\vec D}$ are value assignments to $\pa_{\cal G}(\vec{D}) \setminus \vec{D}$ consistent with $\vec{a}$.

It is then the case that $p(\vec{Y}(\vec{a}))$ is identified if and only if each term $p(\vec{D} \mid \text{do}(\vec{s}_{\vec D}))$ is identified.
To check identifiability of $p(\vec{D} \mid \text{do}(\vec{s}_{\vec D}))$, we will need to introduce graphs derived from ${\cal G}(\vec{V})$ that contains vertices $\vec{R}$ representing random variables, and \emph{fixed} vertices $\vec{S}$ representing intervened on variables.   Such graphs {${\cal G}(\vec{R},\vec{S})$} are called \emph{conditional ADMGs} or CADMGs, and represent {independence restrictions in} interventional distributions {$p(\vec{R} \mid \text{do}(\vec{s}))$}.  The graph in Fig.~\ref{fig:front-door} (e) discussed in \S \ref{sec:proximal-2} is a {conditional DAG (CDAG) -- a} special case of a CADMG.

Given a CADMG ${\cal G}(\vec{R},\vec{S})$,
we say $R \in \vec{R}$ is \emph{fixable} if there \emph{does not exist} another vertex $W$ with both a directed path from $R$ to $W$ in ${\cal G}(\vec{R},\vec{S})$ (e.g. $W$ is a descendant of $R$ in ${\cal G}(\vec{R},\vec{S})$) and a bidirected path from $R$ to $W$.  Note that aside from the starting vertex $R$, and the ending vertex $W$, these two paths need not share vertices.  Given a fixable vertex $R$, a fixing operator $\phi_R({\cal G}(\vec{R},\vec{S}))$ produces a new CADMG ${\cal G}(\vec{R} \setminus \{ R \},\vec{S}\cup\vec\{ R \})$ obtained from ${\cal G}(\vec{R},\vec{S})$ by removing all edges with arrowheads into $R$. {The fixing operator provides a way of constructing the CADMG ${\cal G}(\vec{R} \setminus \{ R \},\vec{S}\cup\vec\{ R \})$ representing restrictions in $p(\vec{R} \setminus \{ R \} \mid \text{do}(\vec{s} \cup r))$, from the CADMG ${\cal G}(\vec{R},\vec{S})$, representing restrictions in $p(\vec{R} \mid \text{do}(\vec{s}))$.}  As we describe below, $R$ being fixable in ${\cal G}(\vec{R},\vec{S})$
is a graphical representation of $p(\vec{R} \setminus \{ R \} \mid \text{do}(\vec{s} \cup r))$ being identified from $p(\vec{R} \mid \text{do}(\vec{s}))$ in a particular way.

A sequence $\sigma_{\vec{J}} \equiv \langle J_1, J_2, \ldots, J_k \rangle$ of vertices in a set $\vec{J} {\subseteq \vec{R}}$ is said to be \emph{valid} in ${\cal G}(\vec{R},\vec{S})$ if it is either empty, or $J_1$ is fixable in ${\cal G}(\vec{R},\vec{S})$, and $\tau(\sigma_{\vec{J}}) \equiv \langle J_2, \ldots, J_k \rangle$ (the \emph{tail} of the sequence) is valid in $\phi_{J_1}({\cal G}(\vec{R},\vec{S}))$.  Any two distinct sequences $\sigma^1_{\vec{J}},\sigma^2_{\vec{J}}$ on the same set $\vec{J}$ valid in ${\cal G}(\vec{R},\vec{S})$ yield the same graph: $\phi_{\sigma^1_{\vec{J}}}({\cal G}(\vec{R},\vec{S})) = \phi_{\sigma^2_{\vec{J}}}({\cal G}(\vec{R},\vec{S}))$.  We will thus write $\phi_{\vec{J}}({\cal G}(\vec{R},\vec{S}))$ to denote this graph.

Each term $p(\vec{D} \mid \text{do}(\vec{s}_{\vec D}))$ in (\ref{eqn:id}) is identified from $p(\vec{V})$ if and only if there exists a sequence $\langle J_1, \ldots \rangle$ of elements in $\vec{J} \equiv \vec{V} \setminus \vec{D}$ valid in ${\cal G}(\vec{V})$.
{
While it might appear that checking whether elements of a set $\vec{J}$ admit a valid sequence in ${\cal G}(\vec{V})$, and finding such a sequence, if it exists, might be a computationally challenging problem, in fact this problem admits a low order polynomial time algorithm.  This is because the fixing operator applied to graphs \emph{only removes edges} and never adds edges.  Removing an edge can never prevent a vertex from being fixable if it was fixable before the edge was removed.  As a result, an algorithm looking for a valid sequence never needs to backtrack -- finding any fixable vertex among vertices yet to be fixed suffices to eventually yield a fixing sequence if it exists, while being unable to find such a vertex at some point implies such a sequence does not exist.
}

If 
a {valid} sequence exists {in ${\cal G}(\vec{V})$ for the set $\vec{J} = \{ J_1, J_2, \ldots \}$}, it implies 
{a set of} identifying assumptions {in the  causal model represented by ${\cal G}(\vec{V})$}.
Let $\vec{R}_1 = \de_{{\cal G}(\vec{V})}(J_1) \setminus \{ J_1 \}$, and $\vec{T}_1 = \vec{V} \setminus (\vec{R}_1 \cup \{ J_1 \})$.
Similarly, let
$\vec{R}_k = \de_{\phi_{\langle J_1, \ldots, J_{k-1} \rangle}({\cal G}(\vec{V}))}(J_k) \setminus \{ J_k \}$ and $\vec{T}_k = \vec{V} \setminus (\{ J_1, \ldots, J_{k-1}, J_k \} \cup \vec{R}_k)$, where $\de_{\cal G}(R_k)$ is the set of descendants of $R_k$ (including $R_k$ itself by convention) in ${\cal G}$.
{Then the graphical causal model implies the following restrictions.}
\begin{assumption}[sequential fixing ignorability]
\label{a:fix-ignore}
{\small
\begin{align}
\label{eqn:ignore-p-1}
\vec{R}_1(j_1) \ci J_1 \mid \vec{T}_1\text{ for all }j_1\\
\label{eqn:ignore-p-2}
\vec{R}_k(j_1,\ldots,j_k) \ci J_k(j_1, \ldots,j_{k-1}) \mid \vec{T}_k(j_1,\ldots,j_{k-1})\text{ for all }j_1,\ldots,j_k.
\end{align}
}
\end{assumption}
{
The identifying assumptions associated with a valid sequence of $\vec{J}$ may be viewed as a{n inductive} generalization of conditional ignorability, or sequential ignorability \citep{robins86new}.
At the point of the induction when a particular variable $J_k(j_1, \ldots, j_{k-1})$ is fixed, 
it is viewed as a ``treatment,'' while all variables in $\vec{R}_k(j_1, \ldots, j_{k-1})$ are viewed as ``outcomes,'' and all variables in $\vec{T}_k(j_1, \ldots, j_{k-1})$ are viewed as ``observed covariates.''
}

Given 
{assumption \ref{a:fix-ignore}}, we obtain identification of $p(\vec{D} \mid \text{do}(\vec{s}_{\vec D}))$ by the following inductive formula:
{\small
\begin{align}
\notag
p(\vec{V} \setminus \{ J_1 \} \mid \text{do}(j_1))
&=
\frac{
p(\vec{V} \setminus \{ J_1 \}, j_1)
}{
p(j_1 \mid \vec{T}_1)
}
=
\frac{
p(\vec{V} \setminus \{ J_1 \}, j_1)
}{
p(j_1 \mid \mb^*_{{\cal G}(\vec{V})}(J_1))
}\\
\notag
p(\vec{V} \setminus \{ J_1, \ldots, J_k \} \mid \text{do}(j_1, \ldots, j_{k}))
&=
\frac{
p(\vec{V} \setminus \{ J_1, \ldots, J_{k} \}, j_k \mid \text{do}(j_1, \ldots, j_{k-1}))
}{
p(j_k \mid \vec{T}_k, \text{do}(j_1, \ldots, j_{k-1}))
}\\
&=
\frac{
p(\vec{V} \setminus \{ J_1, \ldots, J_{k} \}, j_k \mid \text{do}(j_1, \ldots, j_{k-1}))
}{
p(j_k \mid \mb^*_{\phi_{\langle J_1, \ldots, J_{k-1} \rangle}({\cal G}(\vec{V}))}(J_k), \text{do}(j_1, \ldots, j_{k-1}))
},
\label{eqn:one-district}
\end{align}
}
where for any $J_i \in {\vec{V}} \setminus \{ J_1, \ldots, J_{i-1} \}$, $\mb^*_{{ {\cal G}^* }}(J_i)$ denotes all random vertices that are either parents of $J_i$, or that are connected to $J_i$ via collider paths (paths where all consecutive triplets have arrowheads meeting at the middle vertex) {in a CADMG ${\cal G}^*$}.
The operations on the right hand side of (\ref{eqn:one-district}) may be viewed as distributional analogues of the graphical fixing operation $\phi$,
{and are licensed by repeated applications of assumption \ref{a:fix-ignore} and consistency, or alternatively as applications of rule 2 of the potential outcomes calculus \citep{malinsky19po,rrs22volume_id}.
}

As a simple example, we illustrate how identifiability of 
{$p(Y(a))$} in Fig.~\ref{fig:front-door} (b) given 
{by the front-door formula} (\ref{eqn:front-door}) may be reformulated in terms of (\ref{eqn:id}) and (\ref{eqn:one-district}).
The latent projection ADMG of the hidden variable DAG in Fig.~\ref{fig:front-door} (b), sometimes called the \emph{front-door graph}, is shown in Fig.~\ref{fig:front-door} (c).
If we aim to identify $p(Y(a))$ in Fig.~\ref{fig:front-door} (c), we note that $Y^* = \{ Y, M, C \}$, with districts in ${\cal G}_{Y^*}$ being $\{ Y \}, \{ M \}, \{ C \}$.
In fact, valid sequences exists for all sets of elements outside these districts.  Thus, we have the following derivation for the term $p(C \mid \text{do}(a,y,m))$:
{\small
\begin{align*}
p(C,A,M\mid\text{do}(y)) &= \frac{p(C,A,M,y)}{p(y \mid C,A,M)} {= p(C,A,M)}\\
p(C,A\mid\text{do}(y,m)) &= \frac{p(C,A,m \mid \text{do}(y))}{p(m \mid A,C,\text{do}(y))} = \frac{p(C,A,m)}{p(m \mid A,C)} {=p(C,A)}\\
p(C \mid \text{do}(a,y,m)) &= \frac{p(C,a \mid \text{do}(y,m))}{p(a \mid C,\text{do}(y,m))} = \frac{p(C,a)}{p(a \mid C)} = p(C),
\end{align*}
}
the following derivation for the term $p(M \mid \text{do}(a,y,c))$:
{\small
\begin{align*}
p(C,A,M\mid\text{do}(y)) &= \frac{p(C,A,M,y)}{p(y \mid C,A,M)} {= p(C,A,M)}\\
p(C,M\mid\text{do}(y,a)) &= \frac{p(C,a,M\mid\text{do}(y))}{p(a \mid C,\text{do}(y))} = \frac{p(C,a,M)}{p(a \mid C)} {=p(M \mid a,C) p(C)}\\
p(M\mid\text{do}(y,a,c)) &= \frac{p(c,M\mid\text{do}(y,a))}{p(c \mid \text{do}(y,a))} = \frac{p(M \mid a,c) p(c)}{p(c)} = p(M \mid a,c),
\end{align*}
}
and the following derivation for the term $p(Y \mid \text{do}(a,m,c))$:
{\small
\begin{align*}
p(A,M,Y\mid\text{do}(c)) &= \frac{p(c,A,M,Y)}{p(c)} {=p(A,M,Y \mid c)}\\
p(A,Y\mid\text{do}(c,m)) &= \frac{p(A,m,Y\mid\text{do}(c))}{p(m \mid A,\text{do}(c))} = \frac{p(A,m,Y\mid c)}{p(m \mid A,c)} {= p(Y \mid m,A,c) p(A \mid c)}\\
p(Y \mid \text{do}(c,m,a)) &= \frac{p(a,Y\mid\text{do}(c,m))}{p(a\mid Y,\text{do}(c,m))} = \frac{p(Y \mid m,a,c) p(a \mid c)}{ \frac{ p(Y \mid m,a,c) p(a \mid c) }{ \sum_{\tilde{a}} p(Y \mid m,\tilde{a},c) p(\tilde{a} \mid c) } } =
\sum_{\tilde{a}} p(Y \mid m,\tilde{a},c) p(\tilde{a} \mid c).
\end{align*}
}
We then conclude that $p(Y(a))$ is identified from $p(C,A,M,Y)$ via (\ref{eqn:id}) and (\ref{eqn:one-district}) by
{\small
\begin{align*}
& \sum_{m,c} p(Y \mid \text{do}(a,m,c)) p(m \mid \text{do}(a,y,c)) p(c \mid \text{do}(a,y,m))
\\
=&
\sum_{m,c} \left( \sum_{\tilde{a}} p(Y \mid m,\tilde{a},c) p(\tilde{a} \mid c) \right) p(m \mid a, c) p(c),
\end{align*}
}
{which recovers (\ref{eqn:front-door}).}

\begin{figure}[!t]
\centering
\begin{tikzpicture}[>=stealth, node distance=1.2cm]
    \tikzstyle{format} = [draw, very thick, ellipse,
                          minimum size=0.2cm, inner sep=1pt]
    \tikzstyle{unobs} = [draw, very thick, red, ellipse,
                         minimum size=0.2cm, inner sep=1pt]
    \tikzstyle{square} = [draw, very thick, rectangle,
                         minimum size=0.2cm, inner sep=2pt]

    \begin{scope}
        \path[->, very thick]
            node[format] (C) {$C$}
            node[format, right of=C] (M) {$M$}
            node[format, right of=M] (A) {$A$}
            node[format, right of=A] (Y) {$Y$}

	   node[unobs, below of=M, yshift=+0.4cm] (H) {$H$}
	   node[unobs, below of=A, yshift=+0.4cm] (L) {$L$}
	   node[unobs, above of=M, xshift=0.6cm, yshift=-0.4cm] (U) {$U$}

	(C) edge[blue] (M)
	(M) edge[blue] (A)
	(A) edge[blue] (Y)
	(H) edge[red] (C)
	(H) edge[red] (A)
	(L) edge[red] (C)
	(L) edge[red] (Y)
	
	(U) edge[red] (M)
	(U) edge[red] (A)

            node[below of=M, xshift=0.6cm, yshift=0.0cm] (l) {$(a)$}
        ;
    \end{scope}

    \begin{scope}[xshift=4.6cm]
        \path[->, very thick]
            node[format] (C) {$C$}
            node[format, right of=C] (M) {$M$}
            node[format, right of=M] (A) {$A$}
            node[format, right of=A] (Y) {$Y$}


	(C) edge[blue] (M)
	(M) edge[blue] (A)
	(A) edge[blue] (Y)

	

	(M) edge[red, <->, bend left] (A)
	(C) edge[red, <->, bend right=25] (A)
	(C) edge[red, <->, bend right=42] (Y)

            node[below of=M, xshift=0.6cm, yshift=0.0cm] (l) {$(b)$}
        ;
    \end{scope}

    \begin{scope}[xshift=9.2cm]
        \path[->, very thick]
            node[format] (C) {$C$}
            node[format, right of=C] (M) {$M$}
            node[format, right of=M] (A) {$A$}
            node[format, right of=A] (Y) {$Y$}

	   node[format, above of=M, xshift=0.6cm, yshift=-0.4cm] (U) {$U$}


	(C) edge[blue] (M)
	(M) edge[blue] (A)
	(A) edge[blue] (Y)
	
	(U) edge[blue] (M)
	(U) edge[blue] (A)

	(C) edge[red, <->, bend right=25] (A)
	(C) edge[red, <->, bend right=42] (Y)
	

            node[below of=M, xshift=0.6cm, yshift=0.0cm] (l) {$(c)$}
        ;
    \end{scope}

    \begin{scope}[yshift=-3.0cm, xshift=0.0cm]
        \path[->, very thick]
            node[format] (C) {$C$}
            node[format, right of=C] (M) {$M$}
            node[format, right of=M] (A) {$A$}
            node[format, right of=A] (Y) {$Y$}

	   node[unobs, above of=M, xshift=0.6cm, yshift=-0.4cm] (U) {$U$}

	node[format, left of=U] (D) {$D$}
	node[format, left of=D] (Z) {$Z$}
	node[format, right of=U] (W) {$W$}
	node[format, right of=W] (X) {$X$}

	(C) edge[blue] (M)
	(M) edge[blue] (A)
	(A) edge[blue] (Y)
	
	(U) edge[red] (M)
	(U) edge[red] (A)

	(C) edge[red, <->, bend right=25] (A)
	(C) edge[red, <->, bend right=42] (Y)
	
	(W) edge[blue] (A)
	(W) edge[blue] (Y)
	(Z) edge[blue] (M)
	(U) edge[red] (D)
	(U) edge[red] (W)

	(U) edge[red, bend left] (X)
	(W) edge[blue] (X)
	(X) edge[blue] (Y)
	
	(U) edge[red, bend right] (Z)
	(D) edge[blue] (M)
	(D) edge[blue] (A)

	(C) edge[blue] (Z)
	(D) edge[blue] (Z)

            node[below of=M, xshift=0.6cm, yshift=0.0cm] (l) {$(d)$}
        ;
    \end{scope}

    \begin{scope}[yshift=-3.0cm, xshift=5.6cm]
        \path[->, very thick]
            node[format] (C) {$C$}
            node[square, right of=C] (M) {$m$}
            node[format, right of=M] (A) {$A$}
            node[format, right of=A] (Y) {$Y$}

	   node[unobs, above of=M, xshift=0.6cm, yshift=-0.4cm] (U) {$U$}

	node[format, left of=U] (D) {$D$}
	node[format, left of=D] (Z) {$Z$}
	node[format, red, right of=U] (W) {$W$}
	node[format, right of=W] (X) {$X$}

	(M) edge[blue] (A)
	(A) edge[blue] (Y)
	
	(U) edge[red] (A)

	(C) edge[red, <->, bend right=25] (A)
	(C) edge[red, <->, bend right=42] (Y)
	
	(W) edge[red] (A)
	(W) edge[red] (Y)
	(U) edge[red] (D)
	(U) edge[red] (W)
	
	(U) edge[red, bend left] (X)
	(W) edge[red] (X)
	(X) edge[blue] (Y)
	
	(U) edge[red, bend right] (Z)
	(D) edge[blue] (A)

	(C) edge[blue] (Z)
	(D) edge[blue] (Z)

            node[below of=M, xshift=0.6cm, yshift=0.0cm] (l) {$(e)$}
        ;
    \end{scope}

    \begin{scope}[yshift=-3.0cm, xshift=10.0cm]
        \path[->, very thick]
            node[] (C) {}
            node[square, right of=C] (M) {$m$}
            node[format, right of=M] (A) {$A$}
            node[format, right of=A] (Y) {$Y$}

	   node[unobs, above of=M, xshift=0.6cm, yshift=-0.4cm] (U) {$U$}

	node[format, left of=U] (D) {$D$}
	node[ left of=D] (Z) {}
	node[format, red, right of=U] (W) {$W$}
	node[format, right of=W] (X) {$X$}

	(M) edge[blue] (A)
	(A) edge[blue] (Y)
	
	(U) edge[red] (A)

	
	(W) edge[red] (A)
	(W) edge[red] (Y)
	(U) edge[red] (D)
	(U) edge[red] (W)
	
	(U) edge[red, bend left] (X)
	(W) edge[red] (X)
	(X) edge[blue] (Y)
	
	(D) edge[blue] (A)


            node[below of=M, xshift=0.6cm, yshift=0.0cm] (l) {$(f)$}
        ;
    \end{scope}
      
\end{tikzpicture}
\caption{
(a) A hidden variable DAG where $p(Y(a))$ is not identified from $p(C,M,A,Y)$.
(b) A latent projection ADMG of the DAG in (a).
(c) A partial latent projection where the variable $U$ permits identification, provided it is observed.
(d) A partial latent projection where the variable $U$ is hidden, but proximal causal inference using variables $Z,D,W$ and $X$ allow identification.
(e) A conditional graph representing the situation where $M$ is intervened to value $m$, identified by using the proxy $W$, prior to using the proxy $X$ to obtain identification by
marginalizing $C$ and intervening on $A$.
}
\label{fig:one-district}
\end{figure}
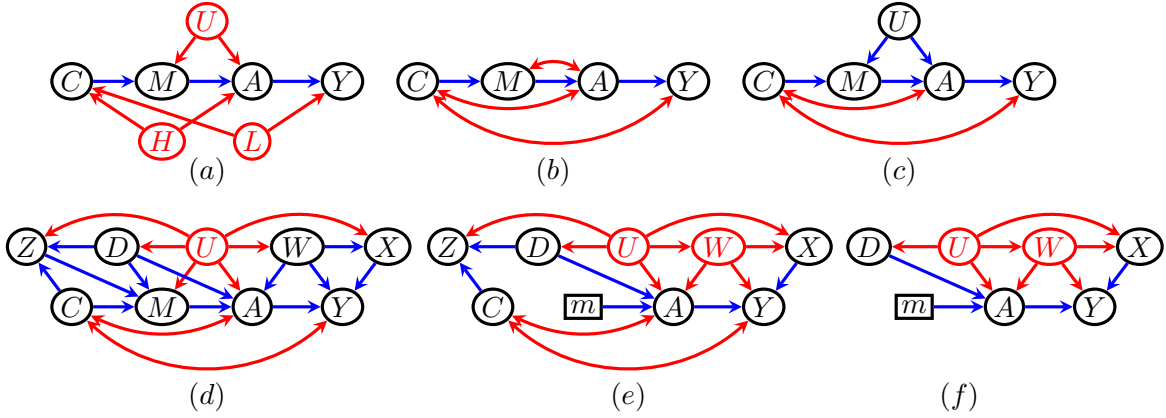

\section{The Proximal ID Algorithm}
\label{sec:new}
The ID algorithm is {sound and} complete for nonparametric identification of interventional distributions $p(\vec{Y}(\vec{a}))$ in causal models represented by any hidden variable DAG
${\cal G}(\vec{V} \cup \vec{H})$, for any disjoint $\vec{Y},\vec{A} \subseteq \vec{V}$.
\emph{Soundness} means that any functional of the observed data $p(\vec{V})$ returned by the ID algorithm for $p(\vec{Y}(\vec{a}))$ is, in fact, equal to $p(\vec{Y}(a))$ in any element in the causal model represented by ${\cal G}(\vec{V})$.
\emph{Completeness} means whenever the ID algorithm fails to return a functional for $p(\vec{Y}(\vec{a}))$, the distribution $p(\vec{Y}(a))$ is, in fact, not identified in the causal model represented by ${\cal G}(\vec{V})$.  In other words, there exist two elements of the hidden variable causal model represented by the DAG ${\cal G}(\vec{V} \cup \vec{H})$ with latent projection ${\cal G}(\vec{V})$, that is two sets of structural equations and noise terms $\{ f^1_{i}(\pa_{\cal G}(V_i),\epsilon^1_i), \epsilon^1_i : V_i \in \vec{V} \cup \vec{H} \}$, $\{ f^2_{i}(\pa_{\cal G}(V_i),\epsilon^2_i), \epsilon^2_i : V_i \in \vec{V} \cup \vec{H} \}$, that yield the same observed marginal distribution $p^1(\vec{V}) = p^2(\vec{V})$, but different interventional distributions of interest $p^1(\vec{Y}(\vec{a})) \neq p^2(\vec{Y}(\vec{a}))$.

Since the ID algorithm is complete, any additional identification results must restrict the causal model in some way in addition to the way the causal model is restricted by ${\cal G}(\vec{V} \cup \vec{H})$.  For example, the average causal effect (ACE) is identified in the instrumental variable model under the assumptions of linearity {of structural equations $f_{i}$}, 
Gaussianity {of noise terms $\epsilon_i$, and correlation of the treatment and instrument variables} \citep{wright1928tariff}, even if the ACE is not identified nonparametrically in this model.  Below, we describe in detail what restrictions are needed in order to obtain additional identification results via proximal causal inference methods that cannot be obtained via the ID algorithm alone.

The distribution $p(\vec{Y}(\vec{a}))$ is not identified {by the ID algorithm} whenever one or more of the terms $p(\vec{D} \mid \text{do}(\vec{s}_{\vec{D}}))$ in (\ref{eqn:id}) cannot be obtained from $p(\vec{V})$.  This, in turn, happens whenever no valid sequence for $\vec{V} \setminus \vec{D}$ exists in ${\cal G}(\vec{V})$.
A variable generally fails to be fixable in the latent projection due to an excessive number of latent variables in the underlying hidden variable DAG.

For example, the distribution $p(Y(a))$ is not identified in the hidden variable causal model shown in Fig.~\ref{fig:one-district} (a).
This is because ${\cal G}_{Y^*}${, which is} obtained from the latent projection of this model shown in Fig.~\ref{fig:one-district} (b), contains a single district $\{ Y \}$, and the set $\vec{V} \setminus \{ Y \} = \{ C, M, A \}$ does not have a valid sequence in Fig.~\ref{fig:one-district} (b).

However, assume we could observe $U$, but not $H$ or $L$.  $p(Y(a))$ would be identified in the implied latent projection graph, shown in Fig.~\ref{fig:one-district} (c), 
since the sequence $\langle M, U, C, A \rangle$ is now valid, yielding:
{\small
\begin{align}
\notag
p(Y,U,C,A\mid\text{do}(m)) &= \frac{p(Y,m,U,C,A)}{p(m \mid U,C)} = p(Y,A \mid m,U,C) p(U,C)\\
\notag
p(Y,C,A\mid\text{do}(m,u)) &= \frac{p(Y,u,C,A\mid\text{do}(m))}{p(u\mid\text{do}(m))} = \frac{p(Y,A\mid m,u,C) p(u,C)}{p(u)}\\
\notag
& = p(Y,A\mid m,u,C) p(C \mid u)\\
\notag
p(Y,A\mid\text{do}(m,u,c)) &= \frac{p(Y,c,A\mid\text{do}(m,u))}{p(c\mid A,Y,\text{do}(m,u))} = \frac{ p(Y,A\mid m,u,c) p(c\mid u)}{\sum_c p(Y,A\mid m,u,c) p(c\mid u)}\\
p(Y\mid\text{do}(m,u,c,a)) &= \frac{p(Y,a\mid\text{do}(m,u,c))}{p(a\mid\text{do}(m,u,c))} = \frac{ \frac{ p(Y,a\mid m,u,c) p(c\mid u)}{\sum_c p(Y,a\mid m,u,c) p(c\mid u)} }{ \sum_y  \frac{ p(y,a\mid m,u,c) p(c\mid u)}{\sum_c p(y,a\mid m,u,c) p(c\mid u)} } = p(Y(a)).
\label{eqn:id-rec}
\end{align}
}
Note that fixing operations above depend on observing $U$.

The ID algorithm, as described in \S \ref{sec:id}, inductively applies a particular identification strategy, where each step may be viewed as 
``adjusting for observed covariates'' $\vec{T}(j_1, \ldots, j_{k-1})$ for the purposes of identifying the distribution of  ``outcomes'' $\vec{R}(j_1, \ldots, j_{k-1})$ after the ``treatment'' $J_k(j_1, \ldots, j_{k-1})$ is intervened on, from the (recursively identified) interventional distribution where variables $J_1, \ldots, J_{k-1}$ have already been intervened on.
Covariate adjustment as an identification strategy may be possible even in the presence of hidden variables.  Specifically, the ID algorithm allows such a strategy provided sequential fixing ignorability holds at every step.  As an example, in the derivation above for the model in Fig.~\ref{fig:one-district} (c), covariate adjustment is valid at each step despite the presence of bidirected edges connecting $C$ and $A$, as well as $C$ and $Y$, representing hidden common causes of those variables.  
Proximal causal inference allows 
a{n alternative} strategy {for dealing with certain hidden variables,} provided appropriate proxies $\vec{Z}$ and $\vec{W}$ 
for {these} hidden variables {exist}, as discussed in \S \ref{subsec:proximal-classic}.

We call hidden variables of the latter type ``proxy-admitting.''
As the previous example shows, the presence of too many hidden variables may render an interventional distribution of interest non-identified, while if even one of these hidden variables were observed, identification may be recovered via the ID algorithm.  What we now show is the requirement that the hidden variable be observed is not always necessary -- provided the hidden variable is ``proxy-admitting,'' instead.  

We illustrate with an example in Fig.~\ref{fig:one-district} (d), where
we will assume the existence of \emph{two} control proxies $Z$ and $D$, and \emph{two} 
{outcome-inducing} proxies $W$ and $X$.
The reason for two sets of proxies is that identification in Fig.~\ref{fig:one-district} (d) will require two applications of covariate adjustment rather than one, and the unobserved variable $U$ must be accounted for in both steps.

We now describe generalizations of assumptions we used in previous examples that allow {identification in Fig.~\ref{fig:one-district} (d).}

We first consider analogues of Assumption \ref{a:proxy-id}.
\begin{assumption}[sequential proxy independence \#1]
\label{a:s-proxy-id-1}
{\small
\begin{align}
\label{eqn:ay-z}
Z \ci A(m), Y(m) \mid U,C,D,X{,M}\\
\label{eqn:w-cond}
W \ci M,Z \mid U,C,D,X
\end{align}
}
\end{assumption}

Next, we assume the analogue of conditional ignorability.
\begin{assumption}[sequential fixing ignorability \#1]
\label{a:s-ignore-1}
{\small
\begin{align}
A(m),Y(m) \ci M \mid U,C,D,X
\label{eqn:ayx-ignore}
\end{align}
}
\end{assumption}
Here the variable $M$ to be fixed is viewed as a treatment, 
descendant variables of $M$, namely $A$ and $Y$, are viewed as outcomes, {and non-descendants of $M$ (excluding the 
{outcome-inducing} proxy $W$), namely $U,C,D,Z,X$, are viewed as covariates.}

The outcome bridge function assumption has the general form of assumption \ref{a:bridge}, but with different treatments, outcomes, and covariates at every step.
Specifically, we have:
\begin{assumption}[sequential outcome bridge function \#1]
\label{a:s-bridge-1}
\begin{itemize}
\item[]
\end{itemize}
There exists a bridge function $b_1(y,a,w,m,c,d,x)$ such that the following equality holds:
{\small
\begin{align}
\label{eqn:integral-y-a-x}
p\left(\underbrace{y,a}_{\text{outcomes}} \mid \underbrace{m,}_{\text{treatment}} \underbrace{z,}_{\text{\stackanchor{control}{proxy}}} \underbrace{c, d, x}_{\text{\stackanchor{observed}{covariates}}}\right) = 
\sum_w b_1(y,a,w,m,c,d,x) p({w} \mid m,z,c,d,x). 
\end{align}
}
\end{assumption}
The completeness assumption \ref{a:complete} is extended slightly for the first step to allow for extra variables in the problem:
\begin{assumption}[sequential completeness \#1]
\label{a:s-complete-1}
\begin{itemize}
\item[]
\end{itemize}
For any random function $v({U})$ in $L^2$,
{\small
\begin{align}
\label{eqn:completeness-2}
\E[v({U}) \mid m,{z},{c},d,x] = 0 \text{ for all }m,{z},c,d \text{ and }{x} \text{ if and only if } v({U}) = 0, \text{ almost surely}.
\end{align}
}
\end{assumption}

As we show {in the proof of the identification result} below, these assumptions, along with positivity imply the following identity:
{\small
\begin{align*}
p(y,a,z,c,d,x \mid \text{do}(m)) = p(Y(m) = y, A(m) = a, z, c, d, x) = \sum_w b_1 p(w,z,c,d,x).
\end{align*}
}

The graph corresponding to the distribution $p(y,a,z,c,d,x \mid \text{do}(m))$ is shown in Fig.~\ref{fig:one-district} (e).  In this distribution, variables $C,Z$ may be safely marginalized without affecting the difficulty of the subsequent subproblem, yielding the distribution $p(y,a,d,x \mid \text{do}(m))$, and the corresponding graph in Fig.~\ref{fig:one-district} (f).
To complete the derivation, analogues of the above assumptions must hold in this distribution.  Specifically, we have:

\begin{assumption}[sequential proxy independence \#2]
\label{a:s-proxy-id-2}
{\small
\begin{align}
\label{eqn:d-y-indep}
D \ci Y(m,a) \mid W, U{,A(m)}\\
\label{eqn:x-a-d-indep}
X \ci A(m),D \mid W, U
\end{align}
}
\end{assumption}

\begin{assumption}[sequential fixing ignorability \#2]
\label{a:s-ignore-2}
{\small
\begin{align}
\label{eqn:y-a-m-ignore}
Y(a,m) \ci A(m) \mid W,U
\end{align}
}
\end{assumption}

\begin{assumption}[sequential outcome bridge function \#2]
\label{a:s-bridge-2}
\begin{itemize}
\item[]
\end{itemize}
There exists a bridge function $b_2(y,x,a)$ such that the following equality holds:
{\small
\begin{align}
p\left(\underbrace{Y(m) = y}_{\text{outcome}} \mid \underbrace{A(m) = a}_{\text{treatment}}, \underbrace{d}_{\text{control proxy}}\right) = \sum_{x} b_2(y, x, a) p(x \mid A(m)=a,d).
\end{align}
}
\end{assumption}

\begin{assumption}[sequential completeness \#2]
\begin{itemize}
\item[]
\end{itemize}
\label{a:s-complete-2}
For any random function $v(W,{U})$ in $L^2$,
{\small
\begin{align}
\label{eqn:completeness-3}
\E_{p^{(m)}}[v(W,U) \mid a,d] = 0 \text{ for all }a\text{ and }d \text{ if and only if }v(W,U) = 0, \text{ almost surely}.
\end{align}
}
\end{assumption}

Finally, the missing directed arrow from $M$ to $Y$ in Fig.~\ref{fig:one-district} (d) represents the following assumption.
\begin{assumption}[exclusion restriction]
\label{a:exclusion-2}
{\small
\begin{align}
Y(a,m) = Y(a).
\label{eqn:exclusion-2}
\end{align}
}
\end{assumption}

Putting everything together, we have the following result.
\begin{theorem}
\label{thm:proximal-example}
Under assumptions \ref{a:s-proxy-id-1}, \ref{a:s-ignore-1}, \ref{a:s-bridge-1}, \ref{a:s-complete-1}, \ref{a:s-proxy-id-2}, \ref{a:s-ignore-2}, \ref{a:s-bridge-2}, \ref{a:s-complete-2}, and positivity,
{\small
\begin{align}
\label{eqn:est-2}
p(Y(a)) &= \sum_{x} b_2(y,x,a) p(x), \text{where }b_2 \text{ is obtained via}\\
\notag
p(Y(m) = y \mid A(m) = a, d) &= \sum_{x} b_2(y, x, a) p(x \mid A(m)=a,d) \text{, which is equivalent to}\\
\notag
\frac{
\sum_{w,x,c} b_1(y,a,x,w,m,c,d) p(w,c,d,x)
}{
\sum_{w,x,c,y} b_1(y,a,x,w,m,c,d) p(w,c,d,x)
} &= \sum_x b_2(y,x,a)
\frac{
\sum_{w,y,c} b_1(y,a,x,w,m,c,d) p(w,c,d,x)
}{
\sum_{w,y,c,x} b_1(y,a,x,w,m,c,d) p(w,c,d,x),
}\\
\notag
&\text{and }b_1 \text{ is obtained via}\\
\notag
p(y,a \mid m, z, c, d, x) &= \sum_w b_1(y,a,x,w,m,d,c) p(w \mid m,c,d,z, x).
\end{align}
}
\end{theorem}

Note that the identifying functional for $p(Y(a))$ is not a function of $m$, although it may appear to depend on $m$ if we examine the syntactic form of the second and third lines of
(\ref{eqn:est-2}).
In particular, we might expect a function $b_2$ that solves the integral equation in the second line to depend on the value of $m$ in the corresponding interventional distributions in that equation.  However, $b_2$ does not, in fact, depend on the value of $m$ under the model in Fig.~\ref{fig:one-district} (d). This is not surprising since assumption \ref{a:exclusion-2} entails the identified interventional distribution $p(Y(a,m)) = p(Y(a))$ is not a function of $m$.  Since the distribution is identified, this restriction manifests as a generalized independence (``Verma'') constraint on the observed data distribution implied by the {causal} model in Fig.~\ref{fig:one-district} (d).

It is instructive to consider the structural similarities of expressions in (\ref{eqn:est-2}) and (\ref{eqn:id-rec}).  The functional in (\ref{eqn:id-rec}) was obtained by fixing $M$, and conditioning on $u$ and $c$, and finally applying the definition of conditioning to the resulting object to obtain $p(Y(m,u,c,a)) {= p(Y(a))}$.  The functional in (\ref{eqn:est-2}) was obtained by fixing $M$ using proximal inference to deal with unobserved confounding between $M$ and its causal descendants, which explains the appearance of the first bridge function $b_1$ in the functional.  Afterwards, variables $Z$ and $C$ are marginalized out, and then subsequent fixing had to be performed using a recursive application of proximal {causal} inference, with another set of proxy variables, and another bridge function $b_2$.  In other words, the fixing sequence that gave rise to (\ref{eqn:id-rec}) is $\langle M,U,C,A \rangle$, while the fixing sequence that gave rise to (\ref{eqn:est-2}) is $\langle M, Z, C, A,X,D \rangle$, with $M$ and $A$ not being fixed in the ``conventional way,'' but with the aid of proxies and bridge functions.  The similarity of the fixing sequences leads to the structural similarity of resulting functionals, just as the proximal g-formula resembles ordinary g-formula, and the proximal front-door functional (\ref{eqn:proximal-frontdoor}) resembles the ordinary front-door functional (\ref{eqn:front-door}).

\subsection{The General Case}
\label{subsec:general-case}

We will now consider general identification theory in hidden variable DAG causal models that combines ideas from the ID algorithm and proximal inference, and that generalizes all examples discussed in the previous sections.  We appropriately name the resulting algorithm the \emph{proximal ID algorithm}.

We will consider hidden variable DAGs with two types of hidden variables: {``ordinary'' hidden variables, handled by the fixing operator of the ID algorithm, and represented by bidirected edges graphically, and ``proxy-admitting'' hidden variables, handled by proximal causal inference, and represented on the graph explicitly as (red) vertices.}

We will denote 
``ordinary'' hidden variables by $\vec{L}$, 
``proxy-admitting'' hidden variables by $\vec{U}$, and observed variables by $\vec{V}$. 
Given a hidden variable DAG ${\cal G}(\vec{V} \cup \vec{U} \cup \vec{L})$, we will formulate our {algorithm} 
using a hidden variable ADMG ${\cal G}(\vec{V} \cup \vec{U})$ obtained by a latent projection operation applied only to $\vec{L}$.


As shown through our derivations in \S \ref{subsec:proximal-classic} and \S \ref{sec:proximal-2}, proximal inference proceeds using a set of control and outcome-inducing proxies. In these previous examples, inductive steps have been formulated in terms of a single (interventional) distribution. In previous derivations, every time we used an outcome-inducing proxy to solve an integral equation to identify the subsequent interventional distribution, we marginalized that proxy. In general, however, we may wish to reuse a proxy variable in multiple integral equations to derive multiple bridge functions.

The distributions we require to solve an integral equation may, in some cases, only require parts of the interventional distribution at the current step of the algorithm. If so, we may only need to identify those parts of the interventional distribution, which may allow us to retain outcome-inducing proxies for subsequent steps. This observation results in some differences in how the algorithm is formulated, compared to the standard ID algorithm.

First, our formulation of the proximal ID algorithm will use a subset $\vec{M} \subseteq \vec{V}$ as 
outcome-inducing proxies, and treat them \emph{as if they were unobserved} when defining the district factorization used in the algorithm.

Given an ADMG ${\cal G}(\vec{V} \cup \vec{U})$, and disjoint $\vec{A},\vec{Y} \subseteq \vec{V}$, fix $\vec{M} \subseteq \vec{V} \setminus (\vec{Y} \cup \vec{A})$, and let $\vec{V}^* \equiv \vec{V} \setminus \vec{M}$.
We start with the following factorization:
{\small
\begin{align}
p(\vec{Y}(\vec{a})) = \sum_{\vec{Y}^* \setminus \vec{Y}} \prod_{\vec{D} \in {\cal D}({\cal G}(\vec{V}^*)_{\vec{Y}^*})} p(\vec{D} \mid \text{do}(\vec{s}_{\vec{D}})),
\label{eqn:proximal-id}
\end{align}
}
where $\vec{Y}^*$ is the set of ancestors of $\vec{Y}$ in ${\cal G}(\vec{V}^*)$ via directed paths that do not intersect $\vec{A}$, and $\vec{s}_{\vec D}$ are value assignments to $\pa_{\cal G}(\vec{D}) \setminus \vec{D}$ consistent with $\vec{a}$.  Note that the districts in the above factorization are defined with respect to a subgraph of ${\cal G}(\vec{V}^*)$, where both $\vec{U}$ and $\vec{M}$ are ``projected out.''

To obtain identification, we must ensure that each term $p(\vec{D} \mid \text{do}(\vec{s}_{\vec{D}}))$ is identified from $p(\vec{V})$ by a combination of regular fixing operations and proximal {causal} inference steps.  To this end, we formulate a set of conditions where a particular treatment variable $A \in \vec{V}$ may be fixed either in the usual way as done by the ID algorithm, or by
{taking advantage of additional assumptions on proxy variables}.  Both kinds of steps are formulated in
hidden variable conditional ADMG{s} (CADMG{s}) ${\cal G}(\vec{V} \cup \vec{M} \cup \vec{U}, \vec{W})$, where $\vec{W}$ represent variables previously fixed, $\vec{U}$ represents unobserved variables that remain relevant in the problem (e.g. have more than one child in $\vec{V} \cup \vec{M}$), $\vec{M}$ represents 
{outcome-inducing} proxies available for use, and $\vec{V}$ represents all other observed variables.  
These CADMG{s} will represent corresponding interventional distribution{s} $p(\vec{v},\vec{m},\vec{u} \mid \text{do}(\vec{w}))$. 

{However, it is not necessary for} the algorithm 
to {(inductively) identify} this entire distribution, but only 
certain marginals that suffice for obtaining identification via {integral equations.} 

Specifically, {at every step} there exists a subset $\vec{V}_1 \subseteq \vec{V}$ and (possibly overlapping) $\vec{M}_1,\vec{M}_2 \subseteq \vec{M}$ such that $\vec{M}_1 \cup \vec{M}_2 = \vec{M}$, and the algorithm has access to $p(\vec{v}_1, \vec{m}_1 \mid \text{do}(\vec{w}))$, and $p(\vec{v}, \vec{m}_2 \mid \text{do}(\vec{w}))$.  We can view the set $\vec{M}_1$ as those proxy variables that have possibly {already} been used for proximal causal inference steps, but remain available for setting up subsequent integral equations, provided only variables in $\vec{V}_1$ are 
{mentioned in these equations}.  We call the distribution $p(\vec{v}_1, \vec{m}_1 \mid \text{do}(\vec{w}))$ the \emph{reusing margin}, {since the proxy variables in this margin are reused by multiple applications of proximal causal inference.}
Variables in $\vec{M}_2$ are proxy variables that have not previously been used, and we call the distribution $p(\vec{v}, \vec{m}_2 \mid \text{do}(\vec{w}))$ containing remaining observed variables, along with proxies $\vec{M}_2$ the \emph{inductive margin}, {since the identification of this margin is obtained by induction from previous such margins.}
During initial steps of the algorithm, both inductive and reusing margins are initialized as $p(\vec{v} \cup \vec{m})$.
We now describe the inductive operation of the {proximal ID} algorithm 
{which involves} both the {steps involving the} 
fixing operation, and {the steps involving} 
proximal {causal} inference. 


\subsubsection{The Proximal ID Algorithm: The Fixing Step}
\label{sssec:fixing}

Suppose ${\cal G}(\vec{V} \cup \vec{M}_2, \vec{W})$ is a latent projection of ${\cal G}(\vec{V} \cup \vec{M} \cup \vec{U}, \vec{W})$.
For $A \in \vec{V}$, if $A$ is fixable in ${\cal G}(\vec{V} \cup \vec{M}_2, \vec{W})$, 
then  it is also fixable in ${\cal G}(\vec{V} \cup \vec{M} \cup {\vec{U}}, \vec{W})$ \citep{richardson17nested}.
This allows us to obtain a new CADMG $\tilde{\cal G}((\vec{V} \cup \vec{M} \cup {\vec{U}}) \setminus \{ A \}, \vec{W} \cup \{ A \}) = \phi_A({\cal G}(\vec{V} \cup \vec{M} \cup {\vec{U}}, \vec{W}))$.
The new distribution corresponding to this CADMG is $p(\vec{v}\setminus\{a\},\vec{m},\vec{u} \mid \text{do}(\vec{w},a))$, where the inductive margin $p(\vec{v}\setminus\{a\},\vec{m}_2 \mid \text{do}(\vec{w},a))$
{may be obtained via}
the following identity, mirroring (\ref{eqn:one-district}):\footnote{In this identity, as before, $\mb^*_{{\cal G}(\vec{V}\cup\vec{M}_2,\vec{W})}(A)$ denotes all random vertices that are either parents of $A$, or that are connected to $A$ via collider paths -- paths where all consecutive triplets have arrowheads meeting at the middle vertex.}
{\small
\begin{align}
p(\vec{V}\setminus \{ A \},\vec{M}_2 \mid \text{do}(\vec{w},a))
&=
\frac{
p(\vec{V} \setminus \{ A \}, \vec{M}_2,a \mid \text{do}(\vec{w}))
}{
p(a \mid \mb^*_{{\cal G}(\vec{V}\cup\vec{M}_2,\vec{W})}(A), \text{do}(\vec{w}))
},
\label{eqn:classic-fix}
\end{align}
}
where the identity is licensed by assumption \ref{a:fix-ignore}, implied by the model for this step since $A$ is fixable in 
{${\cal G}(\vec{V} \cup \vec{M}_2, \vec{W})$}.

In addition,
if $A \in \vec{V}_1$ and $A$ is fixable in ${\cal G}(\vec{V}_1 \cup \vec{M}_1, \vec{W})$, we obtain the reusing margin $p(\vec{v}_1 \setminus \{ a \}, \vec{m}_1 \mid \text{do}(\vec{w},a))$ for subsequent steps, {as follows:}
{\small
\begin{align}
p(\vec{V}_1 \setminus \{ A \}, \vec{M}_1 \mid \text{do}(\vec{w},a))
&=
\frac{
p(\vec{V}_1 \setminus \{ A \}, \vec{M}_1,a \mid \text{do}(\vec{w}))
}{
p(a \mid \mb^*_{{\cal G}(\vec{V}_1\cup\vec{M}_1,\vec{W})}(A), \text{do}(\vec{w}))
}.
\label{eqn:classic-fix-reuse}
\end{align}
}
This identity is, again, licensed by assumption \ref{a:fix-ignore}.

If $A \not\in \vec{V}_1$ or $A$ is not fixable in ${\cal G}(\vec{V}_1 \cup \vec{M}_1, \vec{W})$, we obtain the reusing margin $p(\vec{v}_1', \vec{m}_1' \mid \text{do}(\vec{w},a))$, where
$\vec{V}_1' = \vec{V}_1 \setminus \de_{{\cal G}(\vec{V} \cup \vec{M} \cup \vec{U}, \vec{W})}(A)$, and $\vec{M}_1' = \vec{M}_1 \setminus \de_{{\cal G}(\vec{V} \cup \vec{M} \cup \vec{U}, \vec{W})}(A)$, for subsequent steps, by marginalization:
{\small
\begin{align}
p(\vec{v}_1', \vec{m}_1' \mid \text{do}(\vec{w},a)) = p(\vec{v}_1', \vec{m}_1' \mid \text{do}(\vec{w})) = \sum_{\vec{v}_1 \setminus \vec{v}_1', \vec{m}_1 \setminus \vec{m}_1'} p(\vec{v}_1' ,\vec{m}_1 \mid \text{do}(\vec{w})),
\label{eqn:sum-reuse}
\end{align}
}
provided $\vec{V}_1', \vec{M}_1'$ are non-empty.

\subsubsection{The Proximal ID Algorithm: The Proximal Causal Inference Step}
\label{sssec:proximal}


Given an inductive margin $p(\vec{v}, \vec{m}_2 \mid \text{do}(\vec{w}))$, and a reusing margin $p(\vec{v}_1, \vec{m}_1 \mid \text{do}(\vec{w}))$) identified after an intervention $\text{do}(\vec{w})$, fix $A \in \vec{V}$ where we aim to identify inductive and reusing margins given an intervention $\text{do}(\vec{w},a)$.  To this end we {will need} 
a subset $\vec{M}^* \subseteq \vec{M}$ of 
outcome-inducing proxies (where $A$ is the treatment), and a set of control proxies $\vec{Z} \subseteq \vec{V}$, some of which may be 
outcome-inducing.
Additional conditions must hold for the set $\vec{M}^*$ in order for the bridge function equation to be formulated in terms of available distributions, namely the reusing and inductive margins.  These are described below.

Define $\vec{R} \equiv \de_{{\cal G}(\vec{V} \cup (\vec{M}_2 \setminus \vec{M}^*), \vec{W})}(A) \setminus {(\vec{Z} \cup \{ A \})}$,\footnote{Note that we ``latent project out'' $(\vec{M} \setminus \vec{M}_2) \cup \vec{M}^* \cup \vec{H}$ from the CADMG associated with the current step before defining $\vec{R}$.} and {$\vec{T} \equiv (\vec{V} \cup \vec{M}_2) \setminus (\vec{R} \cup \vec{M}^*)$}.
To apply proximal inference to identify the effect of the intervention on $A$ on the remaining variables in the inductive margin, we assume there exists $\vec{U}^* \subseteq \vec{U}$, 
such that the following set of conditions hold.
{Since these conditions hold in a (recursively identified) distribution where the intervention $\text{do}(\vec{w})$ took place, we index all potential outcomes by $\vec{w}$.}

\begin{assumption}[sequential proxy independence]
\label{a:s-proxy-id-i}
{\small
\begin{align}
\label{eqn:outcome-pre-proxy}
\vec{R}(a,{\vec{w}}) \ci \vec{Z}(a,{\vec{w}}) \mid \vec{U}^*({\vec{w}}),\vec{T}{(
{\vec{w}})} \setminus \vec{Z}(a,{\vec{w}}){,A(\vec{w})}\\
\label{eqn:treatment-post-proxy}
\vec{M}^*({\vec{w}}) \ci A({\vec{w}}),\vec{Z}(a,{\vec{w}}) \mid \vec{U}^*({\vec{w}}), \vec{T}{(
{\vec{w}})} \setminus \vec{Z}{(a,{\vec{w}})}
\end{align}
}
\end{assumption}

\begin{assumption}[sequential fixing ignorability]
\label{a:s-ignore-i}
{\small
\begin{align}
\label{eqn:fix-ignore}
\vec{R}(a,{\vec{w}}) \ci A({\vec{w}}) \mid
\vec{T}{(
{\vec{w}})}\setminus\vec{Z}{(a,{\vec{w}})}
\end{align}
}
\end{assumption}

{A sufficient graphical condition for the assumption \ref{a:s-ignore-i} to hold in $p(\vec{v} \cup \vec{m} \cup \vec{u} \mid \text{do}(\vec{w}))$ is the following condition on the {latent projection ${\cal G}(\vec{V} \cup \vec{U}^*, \vec{W})$ of the} corresponding CADMG ${\cal G}(\vec{V} \cup \vec{M} \cup \vec{U}, \vec{W})$:}
$\de_{{\cal G}(\vec{V} \cup \vec{U}^*, \vec{W})}(A) \cap \dis_{{\cal G}(\vec{V} \cup \vec{U}^*, \vec{W})}(A) = \{ A \}$.

In addition, we impose sequential analogues of the bridge function condition and the completeness condition.

\begin{assumption}[sequential outcome bridge function]
\label{a:s-bridge-i}
\begin{itemize}
\item[]
\end{itemize}
There exists a bridge function $b_A(\vec{m}^*,\vec{r},a,\vec{t}\setminus\vec{z},\vec{w})$ such that the following equality holds:
{\small
\begin{align}
\label{eqn:gen-bridge}
p(\vec{r} \mid a, \vec{t} {\cup \vec{z}}, \text{do}(w)) = \sum_{\vec{m}^*} b_A(\vec{m}^*,\vec{r},a,\vec{t}\setminus\vec{z},\vec{w}) p(\vec{m}^* \mid a, \vec{t} {\cup \vec{z}}, \text{do}(\vec{w}))
\end{align}
}
\end{assumption}
{In order for the integral equation for the bridge function to be exclusively a function of the reusing and/or inductive margins, the only distributions available at the current inductive step, one of two conditions must hold on $\vec{M}^*$.  If $\{ A \} \cup \vec{T} \cup \vec{Z} \subseteq \vec{V}_1$, that is if these variables lie entirely in the reusing margin, then the 
{outcome-inducing} proxies $\vec{M}^*$ must lie either entirely in $\vec{M}_1$ or entirely in $\vec{M}_2$ (that is either entirely in the inductive margin or entirely in the reusing margin).
If $\{ A \} \cup \vec{T} \cup \vec{Z} \not\subseteq \vec{V}_1$, $\vec{M}^*$ must lie entirely in $\vec{M}_2$, that is entirely in the inductive margin.
}

\begin{assumption}[sequential completeness]
\label{a:s-complete-i}
\begin{itemize}
\item[]
\end{itemize}
For any random function $v(U^*)$ in $L^2$,
{\small
\begin{align}
\label{eqn:gen-completeness}
\sum_{\vec{U}^*} v(\vec{U}^*) p(\vec{U}^* \mid a,\vec{t} {\cup \vec{z}},\text{do}(\vec{w})) = 0 \text{ for all }a,\vec{t},{\vec{z},}\vec{w}\text{ if and only if }v(\vec{U}^*) = 0, \text{ almost surely}.
\end{align}
}
\end{assumption}

\begin{theorem}
\label{thm:proxy-fix}
Under assumptions \ref{a:s-proxy-id-i}, \ref{a:s-ignore-i}, \ref{a:s-bridge-i}, \ref{a:s-complete-i}, and positivity,

{\small
\begin{align}
\label{eqn:proxy-fix}
p(\vec{r},\vec{t} \mid \text{do}(\vec{w},a))
&=
\sum_{\vec{u}^*}
\frac{
p(\vec{r},\vec{t},\vec{u}^*,a \mid \text{do}(\vec{w}))
}{
p(a \mid \vec{t},\vec{u}^*,\text{do}(\vec{w}))
}
=
\sum_{\vec{m}^*}
b_A(\vec{m},\vec{r},a,\vec{t} {\setminus \vec{z}},\vec{w})
p(\vec{m}^*,\vec{t} \mid \text{do}(\vec{w}))
\end{align}
}
\end{theorem}

This result yields the inductive margin $p(\vec{r},\vec{t} \mid \text{do}(\vec{w},a))$ for the next step, which potentially includes elements in $\vec{M}_2$ in $\vec{T}$.  To obtain the reusing margin $p(\vec{v}_1, \vec{m}_1 \mid \text{do}(\vec{w}))$ we follow precisely the steps in the previous subsection,
{
via the ordinary fixing operation (\ref{eqn:classic-fix-reuse}), if $A \in \vec{V}_1$ and $A$ is fixable in ${\cal G}(\vec{V}_1 \cup \vec{M}_1, \vec{W})$, or the marginalization operation (\ref{eqn:sum-reuse}) otherwise.}
Note that the inductive margin always marginalizes out the proxies $\vec{M}^*$ that were used in the current step of the algorithm, while the reusing margins aims to keep these proxies in the problem, if possible.
{In addition, only the subset of the control proxies $\vec{Z}$ that are non-descendants of $A$ (in the graph corresponding to the current step), namely $\vec{Z} \cap \vec{T}$, are kept in the inductive margin.  This is necessary since elements in $\vec{Z}$ which are descendants of $A$ are confounded via $\vec{U}^*$, and cannot be treated as a part of $\vec{R}$, since they are being used as control proxies by the algorithm.  Thus, the distribution after $A$ is intervened on that also contains elements in $\vec{Z}$ that are descendants of $A$ would not have been identified.}

The application of proximal causal inference above may be viewed as creating a new subproblem from ${\cal G}(\vec{V} \cup \vec{M}_2 \cup \vec{U},\vec{W})$ by fixing $A$, and viewing
$\vec{M}^*$ as unobserved variables.  We thus extend the fixing operator $\phi(.)$ to apply to ${\cal G}(\vec{V} \cup \vec{M} \cup \vec{U},\vec{W})$, even if $A$ is not fixable
in ${\cal G}(\vec{V} \cup \vec{M}_2, \vec{W})$ as was required in 
\S {\ref{sssec:fixing}}, provided that 
conditions {in \S \ref{sssec:proximal}} are satisfied 
{in} ${\cal G}(\vec{V} \cup \vec{M} \cup \vec{U},\vec{W})$, and the {corresponding} inductive and reusing margins, given $A$ and 
set{s} $\vec{M}^*$,
{$\vec{Z}$ and $\vec{U}^*$.}

To obtain identification of every term $p(\vec{D} \mid \text{do}(\vec{s}_{\vec{D}}))$ in (\ref{eqn:proximal-id}), we need to apply fixing operations and proximal causal inference inductively to the set $\vec{V} \setminus \vec{D}$.  We thus generalize the definition of a valid sequence in the ID algorithm as follows.  Given a CADMG ${\cal G}(\vec{V} \cup \vec{M} \cup {\vec{U}},\vec{W})$, and the corresponding inductive and reusing margins, a sequence $\sigma = \langle J_1, J_2, \ldots \rangle$ is said to be admissible if either $\sigma$ is empty, or $\tau(\sigma)$ is admissible in $\phi_{Z_1}({\cal G}(\vec{V} \cup \vec{M} \cup {\vec{U}},\vec{W}))$, and either {$J_1$} is fixable in ${\cal G}(\vec{V} \cup \vec{M}_2,\vec{W})$, or there {are subsets $\vec{M}^* \subseteq \vec{M},\vec{U}^* \subseteq \vec{U}, \vec{Z} \subseteq \vec{V}$ such that conditions in \S \ref{sssec:proximal} hold, with $J_1$ viewed as a treatment.  In both arms of the disjunction, reusing and inductive margins for subsequent steps of the induction are defined as described above.

Note that unlike the regular fixing operator in \citep{richardson17nested}, which was defined either on CADMGs, or on kernel/CADMG pairs, the notion of an admissible sequence is defined on a triplet of a CADMG, a reusing margin, and an inductive margin.  This is because conditions which allow proximal causal inference to be used at a particular step are not purely graphical, but involve (inductively identified) functions of the observed data distribution as well.
}

We are now ready to state the {main result on the} proximal ID algorithm formally.


\begin{theorem}[proximal ID algorithm]
\label{thm:proximal-id}
Fix an ADMG ${\cal G}(\vec{V} \cup \vec{U})$, with disjoint $\vec{A},\vec{Y} \subseteq \vec{V}$, fix $\vec{M} \subseteq \vec{V} \setminus (\vec{A} \cup \vec{Y})$, $\vec{V}^* \equiv \vec{V} \setminus \vec{M}$, and $\vec{Y}^*$ is the set of ancestors of $\vec{Y}$ in ${\cal G}(\vec{V}^*)$ via directed paths that do not intersect $\vec{A}$.

Then $p(\vec{Y}(\vec{a}))$ is identified from $p(\vec{V})$ in the causal model represented by ${\cal G}(\vec{V} \cup \vec{U})$ given the proxy set $\vec{M}$ if for every $\vec{D} \in {\cal D}({\cal G}(\vec{V}^*)_{\vec{Y}^*})$, the{re exists a} sequence {of elements in the set} $\vec{V}^* \setminus \vec{D}$ 
admissible in ${\cal G}(\vec{V} \cup \vec{U},\vec{W})$.
Furthermore, we then have
{\small
\begin{align}
p(\vec{Y}(\vec{a})) = \sum_{\vec{Y}^* \setminus \vec{Y}} \prod_{\vec{D} \in {\cal D}({\cal G}(\vec{V}^*)_{\vec{Y}^*})} p(\vec{D} \mid \text{do}(\vec{s}_{\vec{D}})),
\label{eqn:proximal-id}
\end{align}
}
with every $p(\vec{D} \mid \text{do}(\vec{s}_{\vec{D}}))$ identified inductively via (\ref{eqn:proxy-fix}) and (\ref{eqn:classic-fix}).
\end{theorem}

We discuss an important special case of the proximal ID algorithm, termed proximal g-computation in \citep{tchetgen2020introduction}, in the Appendix.

\subsection{A Note On Practical Considerations}


The results in the previous sections imply that given any two sequences for a set $\vec{S} \subseteq \vec{V}$ admissible in an ADMG ${\cal G}(\vec{V} \cup \vec{M} \cup \vec{U})$ yield the same object, namely $p(\vec{V} \setminus \vec{S} \mid \text{do}(\vec{s}))$.

Consequently, for any district $\vec{D}$ corresponding to a term $p(\vec{D} \mid \text{do}(\vec{s}_{\vec{D}}))$ appearing in (\ref{eqn:proximal-id}), any two admissible sequences yield the same object, namely that term.  Any such admissible sequence consists of ordinary fixing steps, and steps that resemble fixing to some extent, but use proximal causal inference methods via assumptions described in \S \ref{sssec:proximal}.  As discussed in \S \ref{sec:id}, ordinary fixing steps do not introduce computational complications into the operation of the ID algorithm, since any fixing operation is ``safe'' and does not require the algorithm to backtrack, once performed.  This is also true for the proximal ID algorithm, since any restrictions that hold prior to an ordinary fixing operation being performed will also hold after it is performed.

Thus, the main combinatorial difficulty pertains to choosing sets $\vec{M}^*,\vec{U}^*,\vec{Z}$ necessary for steps involving proximal causal inference.  However, we argue that this difficulty is, to some extent, deceptive.  In causal inference applications, hidden variables suitable for the use of proximal causal inference methods cannot be chosen algorithmically without human insight.  Such variables typically correspond to substantively meaningful confounders that happen, fortuitously, to be associated with appropriate proxies.  Even in cases where a repeated application of proximal causal inference steps is warranted, as in the proximal g-computation algorithm described in the Appendix, every step is substantively justified by appealing to domain specific considerations at that step. That these justifications can be repeated is due to a repeated longitudinal structure of the causal model representing the data.

The appropriate use of the proximal ID algorithm is to clarify whether proximal causal inference methods are appropriate for a particular hidden variable causal model.  This is done by using the district factorization and the fixing operator to subdivide the main problem into a set of subproblems where it is potentially easier for a domain expert to judge whether assumptions underlying proximal causal inference are justified.

Similarly, a hypothetical algorithm that aims to augment the ID algorithm with (nonparametric) methods based on instrumental variables would require human insight in the loop, as any variable that might serve as a candidate instrument if viewed in the causal graph may, in practice, fail to satisfy additional assumptions needed for identification.

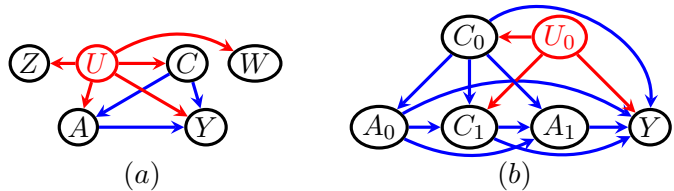
\begin{figure}[!t]
\centering
\begin{tikzpicture}[>=stealth, node distance=1.2cm]
    \tikzstyle{format} = [draw, very thick, ellipse,
                          minimum size=0.2cm, inner sep=1pt]
    \tikzstyle{unobs} = [draw, very thick, red, ellipse,
                         minimum size=0.2cm, inner sep=1pt]
    \tikzstyle{square} = [draw, very thick, rectangle,
                         minimum size=0.2cm, inner sep=2pt]

    \begin{scope}[xshift=0.0cm]
        \path[->, very thick]
            node[format] (A) {$A$}
            node[ above right of=A] (d) {}
            node[format, below right of=d] (Y) {$Y$}

		node[format, xshift=0.6cm, red, left of=d] (U) {${U}$}
		node[format, xshift=-0.6cm, right of=d] (C) {${C}$}

		node[format, xshift=0.3cm, left of=U] (Z) {$Z$}
		node[format, xshift=-0.3cm, right of=C] (W) {$W$}

		(C) edge[blue] (A)
		(C) edge[blue] (Y)
		(A) edge[blue] (Y)
		(U) edge[red] (A)
		(U) edge[red] (Y)
		
		(U) edge[red] (C)
		(U) edge[red, bend left] (W)
		(U) edge[red] (Z)

        node[below of=d, yshift=-0.3cm] (l) {$(a)$}
        ;
    \end{scope}


        \begin{scope}[xshift=4.0cm, node distance=1.2cm]
           \path[->, very thick]
		node[format] (A) {$A_0$}

            node[ above right of=A] (d) {}
		
		node[format, right of=A] (C1) {$C_1$}
		node[format, above of=C1] (C) {$C_0$}

		node[format, right of=C1] (A1) {$A_1$}
		node[format, red, above of=A1] (U) {${U_0}$}
		node[format, right of=A1] (Y) {$Y$}

		(C) edge[blue] (A)
		(A) edge[blue] (C1)
		(A) edge[blue, bend right=25] (A1)
		(A) edge[blue, bend left] (Y)
		
		(C) edge[blue] (C1)
		
		(U) edge[red] (C)
		(U) edge[red] (C1)
		(U) edge[red] (Y)

		(C1) edge[blue] (A1)
		(A1) edge[blue] (Y)
		(C1) edge[blue, bend right=25] (Y)
		
		(C) edge[blue] (A1)
		(C) edge[blue, bend left=65] (Y)

        node[below of=d, xshift=0.95cm, yshift=-0.3cm] (l) {$(b)$}

    	;
        
    \end{scope}
    
\end{tikzpicture}
\caption{
(a) A causal diagram where observed proxy children $Z$ and $W$ of an unobserved confounder $U$ enable proximal inference methods.
(b) A causal diagram corresponding to the sequentially ignorable model, where $p(Y(a_0, a_1))$ is identified via the g-computation algorithm.
}
\label{fig:backdoor-2}
\end{figure}

An important class of hidden variable models where the proximal ID algorithm may be easily applied is one where each of the ``proxy-admitting'' hidden variables $\vec{U}$ has a vector of observed children in the graph, of sufficient cardinality, such that a partition of this vector into control and 
{outcome-inducing} proxies is possible.  An example of a graph representing such a model is shown in Fig.~\ref{fig:backdoor-2}.  In this class of models, the choice of $\vec{M},\vec{U}^*,\vec{Z}$ is trivial, and simply involves choosing any ``proxy-admitting'' hidden variable that prevents the application of regular fixing steps, and partitioning its children proxy variables to obtain $\vec{M}$ and $\vec{Z}$.
Obtaining identification in this class of graphs may be viewed as a generalization of the 
methods presented in \citep{allman2015parameter,kuroki14measurement}.

A particular term in (\ref{eqn:proximal-id}) may be identified by multiple valid sequences of steps, and it may well be the case that some variables are fixed using the standard fixing operator in one sequence, and by the application of proximal causal inference in another.  Choosing among possible sequences when designing an estimator, as well as making other choices that lead to practical estimation strategies introduce highly non-obvious issues that we leave to future work.

%

\subsection{A Note On Characterization of Identification By Proximal Causal Inference Methods}

The ID algorithm yields a total identification result, meaning that whenever it successfully returns an identifying functional for $p(\vec{Y}(\vec{a}))$, this functional equals to the parameter of interest 
{in any element of} the causal model represented by the input graph ${\cal G}$.  This fact allowed prior work \citep{shpitser06id,huang06do} to show that the ID algorithm is \emph{complete}\footnote{Note that this is a different notion of completeness from the completeness assumption discussed earlier, and is instead more closely related to the notion of soundness and completeness of deductive systems.}, meaning that whenever it fails, the target parameter is not identified in the model.

By contrast, the proximal ID algorithm yields a generic identification result, meaning that a function output by the algorithm can only be obtained in a subset of the model represented by the graph where the appropriate Fredholm equations yield solutions.  Historically, it has proven to be very challenging to derive completeness results for generic identification algorithms -- the problem remains open even in the simple case of linear Gaussian causal models.  Thus, the question of whether the proximal ID algorithm is complete is currently open.

\section{Relationship to Identification of Responses to Counterfactual Policies}
\label{sec:dtr}

Consider an ADMG ${\cal G}(\vec{V} \cup \vec{U})$ representing a hidden variable DAG model (where $\vec{U}$ are 
{``proxy-admitting''} hidden variables). Given a set of treatments $\vec{A}$, denote for every $A \in \vec{A}$, a set of non-descendants $\vec{W}_A$ {in ${\cal G}(\vec{V} \cup \vec{U})$}.  Given an outcome $Y$, and a set of known functions (or policies) $\vec{f}_{\vec{A}} \equiv \{ f_A : \mathfrak{X}_{\vec{W}_A} \mapsto \mathfrak{X}_{A} : A \in \vec{A} \}$, {inductively} define $Y(\vec{f}_{\vec{A}})$ as
{\small
\begin{align*}
Y(\{ A \gets f_A(\vec{W}_A(\vec{f}_{\vec{A}})) : A \in \vec{A} \cap \pa_{{\cal G}(\vec{V} \cup \vec{U})}(Y) \}, \{ W(\vec{f}_{\vec{A}}) : W \in \pa_{{\cal G}(\vec{V} \cup \vec{U})}(Y) \setminus \vec{A} \}).
\end{align*}
}
As an example, in Fig.~\ref{fig:backdoor-2} (b), given functions $\vec{f}_{\{A_0, A_1\}} \equiv \{ f_{A_0} : \mathfrak{X}_{C_0} \mapsto \mathfrak{X}_{A_0}; f_{A_1} : \mathfrak{X}_{\{C_0, C_1\}} \mapsto \mathfrak{X}_{A_1} \}$,
{\small
\begin{align*}
Y(\vec{f}_{\{A_0, A_1\!\}} \!\equiv\! Y(C_0, A_0 \!\gets\! f_{A_0}(C_0), C_1(C_0, A_0 \!\gets\! f_{A_0}(C_0)), A_1 \!\gets\! f_{A_1}(C_0, C_1(C_0, A_0 \!\gets\! f_{A_0}(C_0)))).
\end{align*}
}

This counterfactual outcome represents response of $Y$ had values of $A_0$ and $A_1$ been set, possibly contrary to fact, to outputs of functions $f_{A_0}$ and $f_{A_1}$ (which themselves possibly take counterfactual responses to these functions as inputs).  Counterfactuals of this sort arise in precision medicine, analysis of complex longitudinal studies, and reinforcement learning.
{Since most policies of interest deviate from treatment assignment actually observed, the responses are counterfactual, and thus the question of whether they are identified from the observed data distribution, under a given causal model, is crucial.}

Formally, given an ADMG ${\cal G}(\vec{V} \cup \vec{U})$, a set of treatments $\vec{A}$, a subset $\vec{W}_A \subseteq \vec{V} \setminus \vec{A}$ of non-descendants of $A$ for each $A \in \vec{A}$, a set of policies $\vec{f}_{\vec{A}} \equiv \{ f_A : \mathfrak{X}_{\vec{W}_A} \mapsto \mathfrak{X}_{A} : A \in \vec{A} \}$, and a set of outcomes $\vec{Y} \subseteq \vec{V} \setminus \vec{A}$, the question is whether $p(\vec{Y}(\vec{f}_{\vec{A}}))$ is identified from $p(\vec{V})$.

A general identification algorithm for this problem was proposed in \citep{tian08dynamic}, and proven complete (for the unrestricted policy class) in \citep{shpitser18medid}.  In fact, this algorithm reduced the problem of identification of responses to policies to the problem of identification of joint responses to ordinary interventions.  Specifically, $p(\vec{Y}(\vec{f}_{\vec{A}}))$ is identified if and only if $p(\vec{Y}(\vec{a}), \{ \vec{W}_A(\vec{a}) : A \in \vec{A} \})$ is identified.
Moreover, since identification is insensitive to specific values $\vec{a}$ of intervened-on treatments,
the former counterfactual distribution $p(\vec{Y}(\vec{f}_{\vec{A}}))$ may be readily expressed as a marginal of the latter, as follows:
$\sum_{\{ \vec{w}_A : A \in \vec{A} \}} p(\vec{Y}(\{ A \gets f_A(\vec{w}_A) : A \in \vec{A}), \{ \vec{W}_A(\vec{a}) = \vec{w}_A : A \in \vec{A} \})$.
This suggests that the proximal ID algorithm can be readily adapted for identification of responses to policies, essentially without change.

\section{Simulations}
\label{sec:sims}

We now turn to an array of simulation studies to demonstrate how the
identifying assumptions of the proximal ID algorithm can enable unbiased
estimation. We focus on the DAG in Figure \ref{fig:front-door}(d) to highlight how simpler
baseline methods fail when their assumptions are violated.  While the
identification theory allows for $U, C, Z, W$, and $M$ to be vectors, our
simulations is restricted to the univariate case for simplicity, so vector
superscripts are dropped from the notation that follows.
The full details of how we parameterize the synthetic datasets we consider are
in \S~\ref{subsec:synthetic_data}.  Our code implementing our methods and
generating our datasets is included as a supplement.

\subsection{Proximal Front-Door Estimator}

Equation (\ref{eqn:proximal-frontdoor}) provides the identifying functional for the DAG in Figure \ref{fig:front-door}(d).
Our estimator from this functional proceeds in the following steps. First, we
estimate a propensity score model for $M$, $p(M \mid A, Z, C)$.  Using this
model to weight\footnote{We truncate weights at the 2.5th and 97.5th
percentiles.} the observed data distribution gives us the kernel $p^{(m)}$ from
Equation (14), which we use to estimate the bridge function $b^{(m)}(Y \mid w,
a, c, m)$. Following \citet{miao2018confounding}, we estimate this function
using generalized method of moments (GMM).

Once we have the bridge function $b^{(m)}$, we learn a propensity model for
$A$, $p(A \mid Z, C)$ and a regression model for $W$, $E_{(m)}[W \mid A, Z, C]$.
Then, using the observed $Z_i$ and $C_i$ from each row of our observed data
matrix, we sample 100 trajectories of $Y(a=1)$ and $Y(a=0)$ as defined in
Equation (\ref{eqn:proximal-frontdoor}). We then return our estimate of the causal effect as the average over $Y(1) - Y(0)$ for all sampled trajectories and all data rows.

\subsection{Baseline Methods}

To comprehensively benchmark the performance of our proximal front-door estimator,
we compare to several other methods for estimating the causal effect of $A$ on $Y$.

\paragraph{Oracle Baseline.} A motivating assumption for this work is that the unobserved confounder $U$
makes impossible nonparametric identification of the causal effect. If $U$ were in fact observed,
the causal effect could be estimated with a simple application of the g-formula 
which adjusts for the confounding effect of $Z$, $U$, and $C$ \citep{robins86new}.
While our methods are designed for when $U$ is unobserved, comparing against
an oracle with access to $U$ lower-bounds the estimation error for a given
dataset and sample size.

\paragraph{Naive Front-Door.} In Figure \ref{fig:front-door}(d),
the direct effect of $A$ on $Y$ makes the causal effect unidentified by the classic
front-door estimator given by Equation (13).
The front-door estimator relies on the assumption that $(M \ci U \mid A)$, which is
violated in Figure \ref{fig:front-door}(d), so we expect this estimator to return biased estimates
of the causal effect whenever there is a nonzero direct effect of $A$ on $Y$. In the ``{\bf Varied $A \to Y$ Effect}'' experiments below,
we specifically explore how the bias of this estimator changes as we vary the coefficient
responsible for this direct effect.

\paragraph{Simple Proximal.} In Figure \ref{fig:backdoor}(c), the causal effect of $A$ on $Y$ is identified
using proximal inference. In particular, Equation (\ref{eqn:p-g}) provides the identifying functional, which
relies on assumptions (\ref{eqn:assumption-z-i}
 - \ref{eqn:completeness}). In Figure \ref{fig:front-door}(d), assumptions (\ref{eqn:assumption-z-i}) and (\ref{eqn:assumption-w}) are violated by the existence
of $M$ and the path $Z \to M \to W$. Our Simple Proximal baseline uses the identifying functional
from (\ref{eqn:p-g}), ignoring the violation of these assumptions. We expect this estimator to produce
biased estimates of the causal effect whenever there is a nonzero effect along the $Z \to M \to W$ path.
In the ``{\bf Varied $Z \to M \to W$ Effect}'' experiments below, we explore how the bias of this estimator changes as we vary the coefficients
responsible for this path-specific effect. Our implementation of this method draws from the code
released by \citet{miao2018confounding}.

\subsection{Synthetic Data} \label{subsec:synthetic_data}

The choices made in designing a simulation study can heavily influence the
apparent results~\citep{gentzel2019case}. By releasing our code, evaluating over several
randomly-sampled {data generating processes (DGPs)} and datasets, and exploring violations of our method's
assumptions, we strive to make our simulation studies as reproducible and extensible
as possible.

We make several simplifications for our simulation studies. We only consider
univariate $U, C, Z, W,$ and $M$.  We consider settings in which $Z$ and $M$ are
either both binary and or both Gaussian; otherwise, $A$ is always binary and
all other variables are always Gaussian. All effects in the DGP are linear without
interaction terms.

For each of the below experiments, all results are an average of 256 evaluations;
we sample 64 datasets from each of four DGPs.
The coefficients of the four DGPs are randomly sampled from $\text{Unif}(-2, 2)$
and each dataset is sampled with a different random seed.
For the below experiments in which we alter one or two coefficients of the DGP
to explore violations of different assumptions, we do so before sampling the
dataset but leave all other coefficients as is.

For all experiments, we compare each method's estimate of the causal effect
against the true effect as computed analytically from the underlying parameters of the DGP. In the tables below, ``Mean Absolute Bias'' and ``Percent Absolute Bias'' refer to the following computations:
{\small
\begin{align}
\label{eqn:mean_abs_bias}
\text{Mean Absolute Bias} &= \dfrac{1}{N_1}\sum_{i=1}^{N_1}
 \left\vert \dfrac{1}{N_2} \sum_{j=1}^{N_2} (\hat \beta_{i, j} - \beta_{i}) \right\vert \\
\label{eqn:percent_abs_bias}
\text{Percent Absolute Bias} &= \dfrac{1}{N_1}\sum_{i=1}^{N_1}
\dfrac{1}{\vert \beta_i \vert} \left\vert \dfrac{1}{N_2} \sum_{j=1}^{N_2} (\hat \beta_{i, j} - \beta_{i}) \right\vert
\end{align}
}

\noindent
where $N_1=4$ is the number of DGPs and $N_2=64$ is the number of datasets on which we evaluate. $\hat \beta_{i, j}$ is the method's estimate of the causal effect for DGP $i$ and dataset $j$, and $\beta_i$ is the true causal effect for that DGP. Note the absolute value inside the summation over DGPs in (\ref{eqn:mean_abs_bias}) and (\ref{eqn:percent_abs_bias}). This is important because we sample our DGP parameters from $\text{Unif}(-2, 2)$, which is mean zero. Because the effects in our DGP are linear without interactions, the biases of the naive front-door is dependent on the sign and magnitude of the coefficient controlling the $A \to Y$ edge that violates its assumptions. Thus if we average over many DGPs, that coefficient may be approximately mean zero and the bias may disappear, even if the method has high bias for any single DGP. We further explore how specific coefficient values affect the behavior of the different estimators in Tables \ref{tbl:ay_interval}, \ref{tbl:zmw_bias}, and \ref{tbl:uwz_interval}.

\begin{table}[!t]
\centering
\begin{tabular}{l *{6}r}
\toprule
Metric & \multicolumn{3}{c}{Mean Absolute Bias}    &\multicolumn{3}{c}{Percent Absolute Bias} \\
\cmidrule(r){1-1} \cmidrule(lr){2-4} \cmidrule(lr){5-7}
Sample Size          &    4000 &   16000 &   64000 &    4000 &   16000 &   64000 \\
\cmidrule(r){1-1} \cmidrule(lr){2-4} \cmidrule(lr){5-7}
Oracle Backdoor      &   0.007 &   0.003 &   0.002 &   0.007 &   0.003 &   0.001 \\
Naive Front-Door     &   1.003 &   1.005 &   1.003 &   0.790 &   0.791 &   0.789 \\
Simple Proximal      &   1.655 &   1.661 &   1.669 &   1.116 &   1.119 &   1.127 \\
Proximal Front-Door  &   0.006 &   0.009 &   0.004 &   0.005 &   0.008 &   0.003 \\
\bottomrule
\end{tabular}
\caption{{\bf Varied Sample Size} experiments. Each cell of the table shows the metrics in (\ref{eqn:mean_abs_bias}) and (\ref{eqn:percent_abs_bias}) aggregated over four DGPs and 64 datasets per DGP. $Z$ and $M$ are Gaussian.}
\label{tbl:full_bias}
\end{table}

\subsection{Experiments} \label{subsec:synthetic_results}

\paragraph{Varied Sample Size.} We first demonstrate how each method converges
as we increase the sample size. Table~\ref{tbl:full_bias} shows these results
when $Z$ and $M$ are Gaussian. We see both the backdoor oracle and proximal front-door quickly converge towards zero bias.
The proximal front-door estimator achieves bias close to
the oracle, highlighting the effectiveness of proximal methods for recovering
from unobserved confounding.  Neither naive front-door nor the simple proximal
estimators converge towards zero bias as the sample size increases.

\paragraph{Varied $A \to Y$ Effect.} We then examine how the direct effect of $A
\to Y$ empirically introduces bias to the naive front-door estimator. For each
of the four DGPs we consider, we modify the parameters of the sampling distribution
by changing the $A \to Y$ coefficient to a value $\beta_{AY} \in \{0, 0.2, 0.4, 0.8\}$.
For each value $\beta_{AY}$, we sample 256 datasets of 4000 samples.

Table~\ref{tbl:ay_interval}  shows the coverage and interval width of these
methods when we use nonparametric bootstrap with 64 resamplings to produce a
95\% confidence interval. 
We see that when $\beta_{AY} = 0$, the naive front-door estimator's assumptions are met
and it achieves coverage that is comparable with both the oracle method and the proximal front-door, but does so with an interval that is narrower than either. However, as $\beta_{AY}$ increases, the naive front-door estimator coverage quickly drops to 0 while the proximal front-door method maintains the same coverage and similar interval widths. The simple proximal method produces wide intervals with zero coverage for all values of $\beta_{AY}$, as its assumptions do not depend on this direct effect.

\begin{table}[!t]
\centering
\begin{tabular}{l *{8}r}
\toprule
Metric & \multicolumn{4}{c}{Bootstrap Interval Coverage}                  & \multicolumn{4}{c}{Bootstrap Interval Width} \\
\cmidrule(r){1-1} \cmidrule(lr){2-5} \cmidrule(lr){6-9}
$A \to Y$            &       0 &     0.2 &     0.4 &     0.8 &       0 &     0.2 &     0.4 &     0.8 \\
\cmidrule(r){1-1} \cmidrule(lr){2-5} \cmidrule(lr){6-9}
Oracle Backdoor      &  0.938 &   0.938 &   0.938 &   0.938   & 0.338 &   0.338 &   0.338 &   0.338  \\
Naive Front-Door     &  0.766 &   0.219 &   0.020 &   0.000   & 0.258 &   0.254 &   0.251 &   0.245  \\
Simple Proximal      &  0.000 &   0.000 &   0.000 &   0.000   & 0.694 &   0.693 &   0.694 &   0.690  \\
Proximal Front-Door  &  0.945 &   0.941 &   0.945 &   0.941   & 0.595 &   0.595 &   0.595 &   0.597  \\
\bottomrule
\end{tabular}
\caption{{\bf Varied $A \to Y$ Effect} experiments. Width and Coverage of 95\% bootstrap confidence interval as we vary the $A \to Y$ direct effect.
$Z$ and $M$ are Gaussian, we compute intervals with 64 bootstrap resamplings,
and each dataset contains 4000 samples.
The $A \to Y$ direct effect is parameterized by a single coefficient in our linear DGP.
}
\label{tbl:ay_interval}
\end{table}
\begin{table}[!t]
\centering
\begin{tabular}{l *{8}r}
\toprule
Metric              & \multicolumn{4}{c}{Gaussian $Z$, $M$} & \multicolumn{4}{c}{Binary $Z$, $M$} \\
\cmidrule(r){1-1} \cmidrule(lr){2-5} \cmidrule(lr){6-9}
$Z \to M \to W$      &       0 &     0.2 &     0.4 &     0.8 &       0 &     0.2 &     0.4 &     0.8 \\
\cmidrule(r){1-1} \cmidrule(lr){2-5} \cmidrule(lr){6-9}
Oracle Backdoor      &   0.008 &   0.008 &   0.008 &   0.007 & 0.041 &   0.040 &   0.041 &   0.048 \\
Naive Front-Door     &   1.041 &   1.041 &   1.040 &   1.040 & 0.978 &   0.978 &   0.979 &   0.982 \\
Simple Proximal      &   0.012 &   0.319 &   0.985 &   1.404 & 0.065 &   0.081 &   0.104 &   0.208 \\
Proximal Front-Door  &   0.008 &   0.009 &   0.010 &   0.007 & 0.044 &   0.046 &   0.046 &   0.054 \\
\bottomrule
\end{tabular}
\caption{{\bf Varied $Z \to M \to W$ Effect} experiments.
Percent absolute bias as we vary the path-specific effect of $Z \to M \to W$ for either Gaussian or binary $Z$ and $M$.
Each dataset contains 4000 samples.
The effect is parameterized by the $Z \to M$ and $M \to W$ coefficients,
which we set to the same value.}
\label{tbl:zmw_bias}
\end{table}

\paragraph{Varied $Z \to M \to W$ Effect.} When $Z \to M \to W$ path in Figure \ref{fig:front-door}(d) is 0, marginalizing out $M$ gives us the DAG in Figure \ref{fig:backdoor} (c). This explains the low error of the simple proximal estimator in the $0$ columns of Table \ref{tbl:zmw_bias}. However, as soon as the coefficients controlling that path-specific effect increase, assumptions (\ref{eqn:assumption-z-i}) and (\ref{eqn:assumption-w}) -- necessary for the simple proximal estimator's derivation in Equation (\ref{eqn:p-g}) -- are violated and the estimator's bias increases. 

\begin{table}[!t]
\centering
\begin{tabular}{l *{8}r}
\toprule
Metric     & \multicolumn{4}{c}{Bootstrap Interval Coverage} & \multicolumn{4}{c}{Bootstrap Interval Width} \\
\cmidrule(r){1-1} \cmidrule(lr){2-5} \cmidrule(lr){6-9}
$Z \gets U \to W$    &       0 &     0.2 &     0.4 &     0.8 &       0 &     0.2 &     0.4 &     0.8 \\
\cmidrule(r){1-1} \cmidrule(lr){2-5} \cmidrule(lr){6-9}
Oracle Backdoor      &   0.934 &   0.938 &   0.898 &   0.938 &   0.317 &   0.315 &   0.315 &   0.312 \\
Naive Front-Door     &   0.000 &   0.000 &   0.000 &   0.000 &   0.266 &   0.267 &   0.269 &   0.271 \\
Simple Proximal      &   0.242 &   0.070 &   0.043 &   0.000 &   0.492 &   0.538 &   1.467 &   1.411 \\
Proximal Front-Door  &   0.465 &   0.902 &   0.938 &   0.926 &   0.860 &   2.529 &   1.688 &   0.641 \\
\bottomrule
\end{tabular}
\caption{{\bf Varied $Z \gets U \to W$ Effect} experiments. Width and Coverage of 95\% bootstrap confidence interval as we vary
the $Z \gets U \to W$ effects.
$Z$ and $M$ are Gaussian, we compute intervals with 64 bootstrap resamplings,
and each dataset contains 4000 samples.
These effects are parameterized by the $U \to W$ and $U \to Z$ coefficients,
which we set to the same value.}
\label{tbl:uwz_interval}
\end{table}

\paragraph{Varied $Z \gets U \to W$ Effects.} For our final set of simulation studies, we consider the completeness assumptions that require $Z$ and $W$ to be effective proxies for $U$. To explore this assumption, we vary the coefficients that control the direct effect of $U$ on both $Z$ and $W$. When these coefficients are zeroed out, it violates the completeness assumptions given by (\ref{eqn:completeness}) and (\ref{eqn:completeness-q-m}). As these assumptions are essential to learning the bridge functions in (\ref{eqn:fredholm}) and (\ref{eqn:fredholm-q}), we should expect both proximal methods to perform poorly in this setting.

Table \ref{tbl:uwz_interval} shows the width and coverage of a bootstrap confidence interval for each of the four methods as we vary the direct effect of $U$ on its proxies. 
The interval widths in particular provide some empirical evidence for how proximal methods fail when their specific assumptions are violated.
When the coefficients controlling $Z \gets U \to W$ are 0, the bridge function in Equation (\ref{eqn:fredholm-q}) is essentially undefined;
both proximal methods have relatively narrow intervals with poor coverage.
As the coefficient values increase, the proximal front-door quickly increases in coverage to approach that of the oracle.
The simple proximal method does not likewise improve because of the $Z \to M \to W$ path-specific effect that violates assumptions (\ref{eqn:assumption-z-i}) and (\ref{eqn:assumption-w}).

\subsection{Discussion and Limitations}

We have considered a multitude of simulation studies that are designed to comprehensively evaluate the proximal front-door estimator. We show how it generalizes both the front-door estimator and the simple proximal estimator by handling assumption violations of either method. In our results in Tables \ref{tbl:full_bias} and \ref{tbl:zmw_bias}, the proximal front-door estimator has bias comparable to that of the oracle method. In Table \ref{tbl:ay_interval} the proximal front-door interval width is within a factor of $2$ of the oracle while providing the same coverage. These results suggest it can be an empirically effective approach for recovering from unobserved confounding, as long as the completeness assumption explored in Table \ref{tbl:uwz_interval} is met.

Despite these findings, our simulation studies are limited by the simplifying assumptions we have made. Unlike the derivation in Equation (\ref{eqn:proximal-frontdoor}), we consider only univariate $U$, $C$, $Z$, $M$, and $W$. We also only consider linear effects without interaction terms in our DGP. Future work should consider these extensions, but they are not necessary to highlight the empirical behavior of the proximal front-door estimator. We release our code to enable future work, including sampling details necessary to precisely replicate our tables, as an Appendix. 


\section{Analysis Of The Effect Of Methotextrate}
\label{sec:analysis}

We now apply our proximal front-door estimator to an analysis of the effect of
methotextrate (MTX) on tender joint count in patients with rheumatoid
arthritis. The dataset we examine, originally described in
\citet{choi2002methotrexate}, has been studied in several analyses
\citep{fewell2004controlling,whittle2004folate}. Most relevant to this paper are
several recent analyses that have used this data to explore proximal causal
methods \citep{tchetgen2020introduction,ying2021proximal}. While early analyses
of this data have found that MTX reduces mortality
\citep{fewell2004controlling}, recent work proposing causal methods have
highlighted the potential concern for time-varying unobserved confounding
\citep{tchetgen2020introduction}. In particular, while the data contains several
measures of the disease severity and its progression, these variables cannot
capture the full complexity of the patient's underlying health status and
behaviors. Proximal methods allow us to use these noisy measurements as proxies
for the underlying health status that we cannot observe.

Our preprocessing decisions are generally consistent with those of
\citet{tchetgen2020introduction} and \citet{ying2021proximal}; to begin, we
limit our analysis to the 1,010 patients who had at least 12 months of
follow-up. Our outcome $Y$ is the \emph{change} in the patient's tender joint
count after 12 months of follow-up. Our binary treatment is whether the patient
took MTX beginning at baseline.  Our observed covariates $C$ contain nine
categorical variables, including patient age and smoking status. Our two
proxies $Z$ and $W$ are measured at baseline and at seven months of follow-up,
respectively. Each proxy is a vector of four variables, including the
patient's tender joint count and their (categorical) overall assessment of
health.

Our modeling decisions differ from prior work in a number of respects.
Whereas \citet{ying2021proximal} and \citet{tchetgen2020introduction} measured
MTX treatment at two time points, we consider only a single treatment.
The second major difference is the introduction of a mediator $M$, which we define
as a vector of two binary
variables, representing whether the patient was taking Prednisone or
antirheumatic drugs at six months of follow-up. Full details of our
preprocessing and the code to replicate our analyses are included in the
{Appendix}.
Our assumptions are summarized by the DAG in Figure \ref{fig:front-door}(d),
and relax assumptions made by \citet{ying2021proximal} and \citet{tchetgen2020introduction}
in the sense that the 
{outcome-inducing} proxy need no longer be independent of the treatment
conditional on \emph{any} set of variables.  On the other hand, we do assume the treatment is
conditionally ignorable of the mediator variable given covariates.

\begin{table}[!t]
\centering
\begin{tabular}{l *{2}r c *{2}r}
\toprule
& \multicolumn{2}{c}{Point Estimate}  & \multicolumn{1}{c}{Causal Effect} & \multicolumn{2}{c}{Effect Interval} \\
\cmidrule(lr){2-3} \cmidrule(lr){4-4} \cmidrule(lr){5-6}
& $E[Y(0)]$ & $E[Y(1)]$ & $E[Y(1) - Y(0)]$ & $2.5\%$ & $97.5\%$ \\
\cmidrule(r){1-1} \cmidrule(lr){2-3} \cmidrule(lr){4-4} \cmidrule(lr){5-6}
Naive Front-Door     & -0.155  &  -0.137 & 0.017 & -0.026 &   0.063   \\
Simple Proximal      & -0.171  &  -0.908 & -0.737 & -1.144 &   -0.392   \\
Proximal Front-Door  &  0.149  &  -0.193 & -0.343 & -0.977 &   0.247   \\
\bottomrule
\end{tabular}
\caption{Results of three estimators with binary treatment, where $a=1$ only for patients who began MTX at or before baseline. The Point Estimate columns show estimates for $E[Y(a)]$, the expected \emph{change} in tender joint count for a patient counterfactually assigned treatment $a$. The Effect Interval columns show a 95\% confidence interval of the causal effect, $E[Y(1) - Y(0)]$, using 256 bootstrap resamplings. While the point estimates of both proximal methods suggest MTX helps reduce tender joint count, the proximal front-door estimator's higher variance results in an insignificant estimate of the causal effect.
}
\label{tbl:mtx_two_timesteps}
\end{table}

Similar to \S \ref{sec:sims}, we compare our proximal front-door estimator
against a naive front-door estimator that assumes no direct effect $A \to Y$,
and against a simple proximal estimator that assumes no path-specific effect $Z
\to M \to W$.  The simple proximal estimator is most similar to the past work
of \citet{ying2021proximal}. As clinician domain knowledge in prior work
strongly suggests the presence of the $A \to Y$ edge, we might expect the
naive front-door estimator to be biased.

We start with the simplest case where $A$ is binary: either patients received
MTX starting at baseline ($A=1$) or they did not ($A=0$). Table
\ref{tbl:mtx_two_timesteps} shows each method's point estimate for the expected
counterfactual \emph{change} in tender joint count without or with MTX. A negative value
of $E[Y(a)]$ suggests that on average, treatment $a$ reduces patients' tender
joint count.
The naive front-door estimates the causal effect ($E[Y(1) - Y(0)]$) at 0.017, suggesting that both treatments have roughly equal effectiveness.
In contrast, the simple proximal and proximal front-door methods estimate this
causal effect at -0.737 and -0.343 respectively, suggesting that MTX is more effective at reducing tender joint count.

The right-most columns of Table \ref{tbl:mtx_two_timesteps} show a
95\% confidence interval for this causal effect, calculated from 256 bootstrap resamplings.  While we cannot know
the true causal effect, the bootstrap interval width provides insight into the
variance of the estimators. As with our simulation studies, the naive
front-door estimator has the lowest variance, and our proximal front-door
estimator has the highest.

\begin{figure}[!t]
\centering
\includestandalone[width=\textwidth,mode=image]{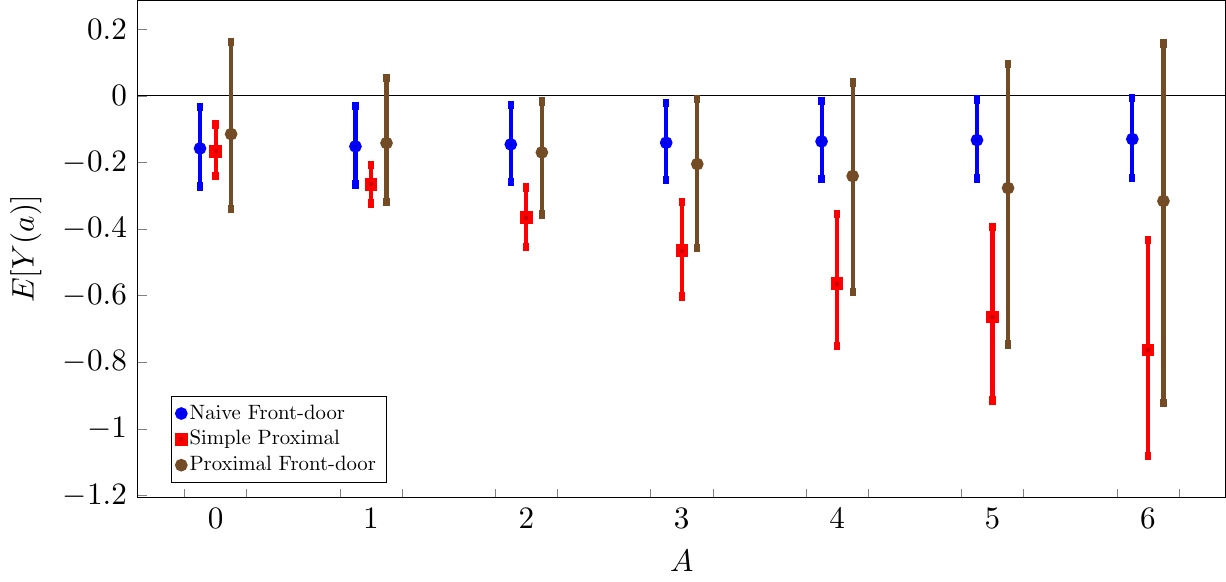}
\caption{Results of three estimators with a categorical treatment ranging from $a=0$ (patients with no MTX) to $a=6$ (patients who began MTX at or before baseline). The figure shows for each estimator and each value of $a$, the point estimate of $E[Y(a)]$ and the 95\% confidence interval from 256 bootstrap resamplings. If we calculate causal effects $E[Y(a) - Y(0)]$ for each estimator, the simple proximal gives significant effects for each $a$, according to a 95\% bootstrap confidence interval. However, neither front-door estimator gives any significant effects.}
\label{fig:mtx_six_timesteps}
\end{figure}

One concern specific to the data and our results in Table
\ref{tbl:mtx_two_timesteps} is that our treated patients are defined as
\emph{only} those who began treatment at or before baseline; all patients who
started treatment after baseline are coded as untreated. We explore
the effect of cumulative exposure by defining treatment as the patient's months on MTX,
rather than a binary indicator. Because we have an edge $A \to M$ in
Figure \ref{fig:front-door} (d) and $M$ is measured at six months of follow-up,
we define $A=0$ for patients who were never treated or were only treated sometime
after six months of follow-up. We define $A=6$ for patients who were treated
at or before baseline. For patients who began treatment at $t=1, \ldots, 5$
months of follow-up, we define $A=6-t$: a patient with five months of
pre-mediator treatment began treatment at one month of follow-up. Our
categorical definition reduces the number of individuals within each strata of
$A$, with 826 patients labeled with $A=0$, 123 patients with $A=6$, and
only 61 total patients with $A$ between 1 and 5. Note that patients who were labeled as $A=1$
in the analysis for Table \ref{tbl:mtx_two_timesteps} are now labeled as $A=6$; we cannot further subdivide these patients because the dataset does not indicate whether they started MTX at baseline or several years beforehand.

Figure \ref{fig:mtx_six_timesteps} shows all three methods with this
categorically-defined $A$. As in the binary case, the simple proximal estimator
suggests that increased MTX reduces tender joint count. For the causal effect
defined as $E[Y(6) - Y(0)]$, the naive front-door method gives an estimate of 0.028, suggesting no benefit from MTX. The simple proximal method gives an effect estimate of -0.596, which is significant according to a 95\% bootstrap confidence interval from 256 resamplings. The proximal front-door estimator gives a more conservative estimate of the causal effect at -0.201, which is not significant.
The 95\% bootstrap confidence intervals for the proximal methods are again
quite wide, especially for larger values of $A$.

Code for preprocessing the data and reproducing these results are provided in the
{Appendix}.

\section{Conclusions}
\label{sec:conclusions}

We have introduced the proximal ID algorithm, a synthesis of proximal causal inference, and nonparametric identification theory based on graphical causal models.

On the one hand, the proximal ID algorithm is able to obtain identification in cases where the classical ID algorithm fails by exploiting additional assumptions involving proxy variables.  On the other hand, the proximal ID algorithm greatly extends the applicability of proximal causal inference methods by taking advantage of graphical identification theory.

{As an important special case of our algorithm, we introduced the proximal generalization of} the front-door criterion \citep{pearl95causal}{, which no longer requires that the mediator capture all causal effects of the treatment on the outcome.}
{In the Appendix, we review another important special case: the} proximal extensions of the g-computation algorithm \citep{robins86new}.
Since estimation methods for the functionals arising in the latter case have already been considered in the literature \citep{tchetgen2020introduction}, we illustrate estimation for the proximal front-door functional by means of simulation studies and a data application.

The validity of the output of the proximal ID algorithm relies on a correctly specified graphical causal model, as well as the existence of proxy variables and completeness conditions that must hold within certain subproblems encountered by the algorithm.

\clearpage
\appendix

\section{Proofs}
\label{appendix:proofs}

\begin{thma}{\ref{thm:proximal-example}}
Under assumptions \ref{a:s-proxy-id-1}, \ref{a:s-ignore-1}, \ref{a:s-bridge-1}, \ref{a:s-complete-1}, \ref{a:s-proxy-id-2}, \ref{a:s-ignore-2}, \ref{a:s-bridge-2}, \ref{a:s-complete-2}, and positivity,
{\small
\begin{align}
\label{eqn:est-3}
p(Y(a)) &= \sum_{x} b_2(y,x,a) p(x), \text{where }b_2 \text{ is obtained via}\\
\notag
p(Y(m) = y \mid A(m) = a, d) &= \sum_{x} b_2(y, x, a) p(x \mid A(m)=a,d) \text{, which is equivalent to}\\
\notag
\frac{
\sum_{w,x,c} b_1(y,a,x,w,m,c,d) p(w,c,d,x)
}{
\sum_{w,x,c,y} b_1(y,a,x,w,m,c,d) p(w,c,d,x)
} &= \sum_x b_2(y,x,a)
\frac{
\sum_{w,y,c} b_1(y,a,x,w,m,c,d) p(w,c,d,x)
}{
\sum_{w,y,c,x} b_1(y,a,x,w,m,c,d) p(w,c,d,x),
}\\
\notag
&\text{and }b_1 \text{ is obtained via}\\
\notag
p(y,a \mid m, z, c, d, x) &= \sum_w b_1(y,a,x,w,m,d,c) p(w \mid m,c,d,z, x).
\end{align}
}
\end{thma}
\begin{proof}
We first obtain the following:
{\small
\begin{align}
(\ref{eqn:ay-z})
& \underbrace{\Rightarrow}_{\text{consistency}}
\label{eqn:z-cond}
A,Y \ci Z \mid M=m,U,C,D,X\\
\notag
{
(\ref{eqn:ay-z})
+
(\ref{eqn:ayx-ignore})
}
&
{
\underbrace{\Rightarrow}_{\text{contraction}}
Z,M \ci A(m), Y(m) \mid U,C,D,X}\\
\label{eqn:ayx-ignore-2}
&
{
\underbrace{\Rightarrow}_{\text{weak union}}
M \ci A(m), Y(m) \mid U,C,D,X,Z
}
\\
\label{eqn:w-cond-2}
(\ref{eqn:w-cond})
& \underbrace{\Rightarrow}_{\text{weak union}} W \ci Z \mid M,U,C,D,X\\
\label{eqn:w-cond-3}
(\ref{eqn:w-cond})
& \underbrace{\Rightarrow}_{\text{weak union}} W \ci M \mid Z,U,C,D,X
\end{align}
}

We then obtain the following derivation, where we will denote $b_1(y,a,x,w,m,d,c)$ by $b_1$ {for conciseness}:
{\small
\begin{align}
\notag
p(y,a \mid m, z, c, d, x) = \sum_w b_1 p(w \mid m,c,d,z, x) \Rightarrow (\text{by }(\ref{eqn:z-cond}),(\ref{eqn:w-cond-2}))\\
\notag
\sum_u p(y,a \mid m, c, d, x, u) p(u \mid m,z,c,d, x) = \sum_w b_1 \sum_u p(w \mid m,c,d,x,u) p(u \mid m,z,c,d,x) \Rightarrow
(\text{by }(\ref{eqn:completeness-2}))\\
\notag
p(y,a \mid m, c, d, x, u) = \sum_w b_1 p(w \mid m,c,d,x,u) \Rightarrow (\text{by }(\ref{eqn:z-cond}),(\ref{eqn:w-cond-2}),(\ref{eqn:w-cond-3}))\\
\notag
\sum_{u}
p(y,a \mid m, c, d, x, u, z) p(c,d,x, u, z) = \sum_{u} \sum_w b_1 p(w \mid c,d,x,u,z) p(c,d,x, u, z) \Rightarrow (\text{by }(\ref{eqn:ayx-ignore-2}))\\
\label{eqn:final-indicator}
p(Y(m) = y, A(m) = a \mid c, d, x, z) p(c, d, x, z) = p(y,a,c,d,x, z \mid \text{do}(m)) = \sum_w b_1 p(w,c,d,x, z).
\end{align}
}

Next, we follow the same derivation structure, but now with the second set of assumptions, corresponding to the second fixing operation we must perform.
{\small
\begin{align}
(\ref{eqn:d-y-indep})
&\underbrace{\Rightarrow}_{\text{consistency}} Y(m) \ci D \mid A(m)=a, W, U.
\label{eqn:y-d-indep}\\
\label{eqn:y-a-m-ignore-2}
{
(\ref{eqn:d-y-indep})
+
(\ref{eqn:y-a-m-ignore})
}
&
{
\underbrace{\Rightarrow}_{\text{contraction}}
D,A(m) \ci Y(m,a) \mid W, U
\underbrace{\Rightarrow}_{\text{weak union}}
A(m) \ci Y(m,a) \mid W, U, D
}
\\
\label{eqn:x-a-d-indep-2}
(\ref{eqn:x-a-d-indep})
& \underbrace{\Rightarrow}_{\text{weak union}}
X \ci D \mid A(m), W, U\\
\label{eqn:x-a-d-indep-3}
(\ref{eqn:x-a-d-indep})
& \underbrace{\Rightarrow}_{\text{weak union}}
X \ci A(m) \mid D, W, U
\end{align}
}

Noting that both $p(Y(m) = y \mid A(m) = a, d)$ and $p(x \mid A(m)=a,d)$ are functions of $p(y,a,c,d,x, z \mid \text{do}(m))$,
which is identified by the above argument, we continue with the derivation using assumption \ref{a:s-bridge-2}, and denoting $b_2(y,x,a)$ by $b_2$ for conciseness, as before:
{\small
\begin{align*}
p(Y(m) = y \mid A(m) = a, d) = \sum_{x} b_2 p(x \mid A(m)=a,d) \Rightarrow \\
\hspace{5.0cm} (\text{by }(\ref{eqn:y-d-indep}),(\ref{eqn:x-a-d-indep-2}))\\
\sum_{w,u} p(Y(m) = y \mid A(m) = a, w,u) p(w,u \mid a, d) = \sum_{x} b_2 \sum_{w,u} p(x \mid A(m)=a,w,u) p(w,u \mid a,d)
\Rightarrow \\
\hspace{5.0cm} (\text{by }(\ref{eqn:completeness-3}))\\
p(Y(m) = y \mid A(m) = a, w,u) = \sum_{x} b_2 p(x \mid A(m)=a,w,u)\Rightarrow (\text{by }(\ref{eqn:y-d-indep}),(\ref{eqn:x-a-d-indep-2}),(\ref{eqn:x-a-d-indep-3})
)\\
\sum_{w,u,d} p(Y(m) = y \mid A(m) = a, w,u,d) p(w,u,d) = \sum_{w,u} \sum_{x,d} b_2 p(x \mid w,u,d) p(w,u,d)\Rightarrow(\text{by }(\ref{eqn:y-a-m-ignore-2}),(\ref{eqn:exclusion-2}))\\
\sum_{w,u,d} p(Y(m,a)=y \mid w,u,d) p(w,u,d) = p(Y(a)) = 
\sum_{x,d} b_2(y,x,a) p(x,d) = \sum_{x} b_2(y,x,a) p(x).
\end{align*}
}
This completes the proof.
\end{proof}

\begin{thma}{\ref{thm:proxy-fix}}
Under assumptions \ref{a:s-proxy-id-i}, \ref{a:s-ignore-i}, \ref{a:s-bridge-i}, \ref{a:s-complete-i}, and positivity,

{\small
\begin{align}
\label{eqn:proxy-fix-2}
p(\vec{r},\vec{t} \mid \text{do}(\vec{w},a))
&=
\sum_{\vec{u}^*}
\frac{
p(\vec{r},\vec{t},\vec{u}^*,a \mid \text{do}(\vec{w}))
}{
p(a \mid \vec{t},\vec{u}^*,\text{do}(\vec{w}))
}
=
\sum_{\vec{m}^*}
b_A(\vec{m},\vec{r},a,\vec{t}{\setminus\vec{z}},\vec{w})
p(\vec{m}^*,\vec{t} \mid \text{do}(\vec{w}))
\end{align}
}
\end{thma}
\begin{proof}
We first obtain the following:
{\small
\begin{align}
(\ref{eqn:outcome-pre-proxy})
\label{eqn:outcome-pre-proxy-obs}
& \underbrace{\Rightarrow}_{\text{consistency}} \vec{R}({\vec{w}}) \ci \vec{Z}({\vec{w}}) \mid A({\vec{w}})=a,\vec{U}^*({\vec{w}}), \vec{T}({\vec{w}}) \setminus \vec{Z}({\vec{w}})\\
\notag
{
(\ref{eqn:outcome-pre-proxy})
+
(\ref{eqn:fix-ignore})
}
&
{
\underbrace{\Rightarrow}_{\text{contraction}}
\vec{R}(a,\vec{w}) \ci \vec{Z}(a,\vec{w}),A(\vec{w}) \mid \vec{U}^*(\vec{w}),\vec{T}{(\vec{w})} \setminus \vec{Z}(a,\vec{w})}
\\
\label{eqn:fix-ignore-2}
&{
\underbrace{\Rightarrow}_{\text{weak union, decomposition}}
\vec{R}(a,\vec{w}) \ci A(\vec{w}) \mid \vec{U}^*(\vec{w}),\vec{T}{(\vec{w})}
}
\\
\notag
(\ref{eqn:treatment-post-proxy})
& \underbrace{\Rightarrow}_{\text{weak union}}
\vec{M}^*({\vec{w}}) \ci \vec{Z}(a{,\vec{w}}) \mid A({\vec{w}}),\vec{U}^*({\vec{w}}), \vec{T}(
{\vec{w}}) \setminus \vec{Z}{(a{,\vec{w}})}\\
\label{eqn:treatment-post-proxy-2}
&\underbrace{\Rightarrow}_{\text{consistency}}
\vec{M}^*({\vec{w}}) \ci \vec{Z}({\vec{w}}) \mid A({\vec{w}})=a,\vec{U}^*({\vec{w}}), \vec{T}({\vec{w}}) \setminus \vec{Z}({\vec{w}})
\\
\label{eqn:treatment-post-proxy-3}
(\ref{eqn:treatment-post-proxy})
& \underbrace{\Rightarrow}_{\text{weak union}}
\vec{M}^*({\vec{w}}) \ci A({\vec{w}}) \mid \vec{Z}(a{,\vec{w}}),\vec{U}^*({\vec{w}}), \vec{T}(
{\vec{w}}) 
\end{align}
}

We then get the following derivation, where we denote $b_A(\vec{m}^{{*}},\vec{r},a,\vec{t}{\setminus\vec{z}},\vec{w})$ by $b_A$ for conciseness:
{
\small
\begin{align}
\notag
p(\vec{r} \mid a,\vec{t}{\cup\vec{z}},\text{do}(\vec{w}))
&= \sum_{\vec{m}^*} b_A p(\vec{m}^* \mid a,\vec{t}{\cup\vec{z}},\text{do}(\vec{w}))
\Rightarrow (\text{by }(\ref{eqn:outcome-pre-proxy-obs}),(\ref{eqn:treatment-post-proxy-2}))\\
\notag
\sum_{\vec{u}^*}
p(\vec{r} | \vec{u}^*\!,a,\vec{t}\setminus\!\vec{z},\text{do}(\vec{w}))
p(\vec{u}^*\! | a,\vec{t} {\cup \vec{z}},\text{do}(w))
&=
\sum_{\vec{m}^*} b_A
\sum_{\vec{u}^*}
p(\vec{m}^*\! | a,\vec{t}\setminus\!\vec{z},\vec{u}^*\!,\text{do}(\vec{w}))
p(\vec{u}^*\! | a,\vec{t}{\cup\vec{z}},\text{do}(\vec{w}))
\Rightarrow \\
\notag
& \hspace{4.0cm} (\text{by }(\ref{eqn:gen-completeness}))\\
\notag
p(\vec{r} \mid \vec{u}^*,a,\vec{t}\setminus\vec{z},\text{do}(\vec{w}))
&=
\sum_{\vec{m}^*} b_A
p(\vec{m}^* \mid a,\vec{t}\setminus\vec{z},\vec{u}^*,\text{do}(\vec{w}))
\Rightarrow \\
\notag
& \hspace{4.0cm} (\text{by }(\ref{eqn:outcome-pre-proxy-obs}),(\ref{eqn:treatment-post-proxy-2}),(\ref{eqn:treatment-post-proxy-3}))\\
\notag
\sum_{\vec{u}^*}
p(\vec{r} \mid \vec{u}^*,a,\vec{t},\text{do}(\vec{w}))
p(\vec{u}^*,\vec{t} \mid \text{do}(\vec{w}))
&=
\sum_{\vec{u}^*}
\sum_{\vec{m}^*} b_A
p(\vec{m}^* \mid \vec{t},\vec{u}^*,\text{do}(\vec{w}))
p(\vec{u}^*,\vec{t} \mid \text{do}(\vec{w}))
\Rightarrow \\
\notag
& \hspace{4.0cm} (\text{by }(\ref{eqn:fix-ignore-2}),\text{ and consistency})\\
\label{eqn:proxy-fix-3}
\sum_{\vec{u}^*}
\frac{
p(\vec{r},\vec{t},\vec{u}^*,a \mid \text{do}(\vec{w}))
}{
p(a \mid \vec{t},\vec{u}^*,\text{do}(\vec{w}))
}
&=
p(\vec{r},\vec{t} \mid \text{do}(\vec{w},a))
=
\sum_{\vec{m}^*} b_A p(\vec{m}^*,\vec{t} \mid \text{do}(\vec{w}))
\end{align}
}
This establishes the result.
\end{proof}

\begin{thma}{\ref{thm:proximal-id}}
{\bf (proximal ID algorithm)}
Fix an ADMG ${\cal G}(\vec{V} \cup \vec{U})$, with disjoint $\vec{A},\vec{Y} \subseteq \vec{V}$, fix $\vec{M} \subseteq \vec{V} \setminus (\vec{A} \cup \vec{Y})$, $\vec{V}^* \equiv \vec{V} \setminus \vec{M}$, and $\vec{Y}^*$ is the set of ancestors of $\vec{Y}$ in ${\cal G}(\vec{V}^*)$ via directed paths that do not intersect $\vec{A}$.

Then $p(\vec{Y}(\vec{a}))$ is identified from $p(\vec{V})$ in the causal model represented by ${\cal G}(\vec{V} \cup \vec{U})$ given the proxy set $\vec{M}$ if for every $\vec{D} \in {\cal D}({\cal G}(\vec{V}^*)_{\vec{Y}^*})$, the{re exists a} sequence {of elements in the set} $\vec{V}^* \setminus \vec{D}$ 
admissible in ${\cal G}(\vec{V} \cup \vec{U},\vec{W})$.
Furthermore, we then have
{\small
\begin{align}
p(\vec{Y}(\vec{a})) = \sum_{\vec{Y}^* \setminus \vec{Y}} \prod_{\vec{D} \in {\cal D}({\cal G}(\vec{V}^*)_{\vec{Y}^*})} p(\vec{D} \mid \text{do}(\vec{s}_{\vec{D}})),
\label{eqn:proximal-id-2}
\end{align}
}
with every $p(\vec{D} \mid \text{do}(\vec{s}_{\vec{D}}))$ identified inductively via (\ref{eqn:proxy-fix}) and (\ref{eqn:classic-fix}).
\end{thma}
\begin{proof}
First, note that standard causal models of a hidden variable DAG ${\cal G}(\vec{V} \cup \vec{H})$ imply that every interventional distribution, including $p(\vec{Y}^*(\vec{a}))$, {admits a district factorization} with respect to the 
induced subgraph ${\cal G}(\vec{V})_{\vec{Y}^*}$ of the latent projection ADMG ${\cal G}(\vec{V}))$ of ${\cal G}(\vec{V} \cup \vec{H})$.
The conclusion then follows inductively (on the fixing sequence of every term) from the soundness of the ID algorithm (see e.g. the proof in \citep{richardson17nested}), and the derivation yielding (\ref{eqn:proxy-fix}).
\end{proof}

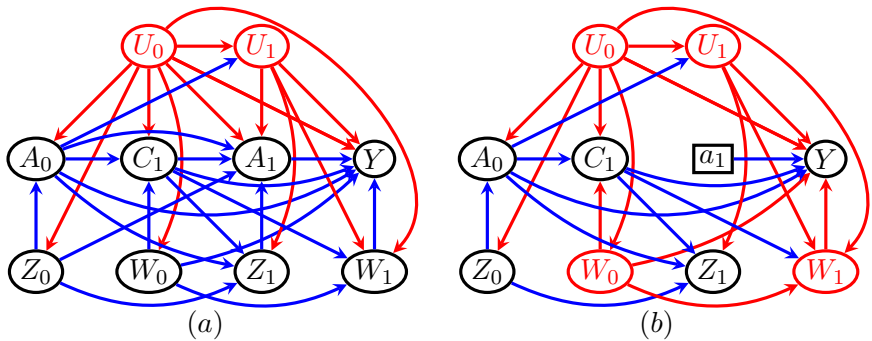
\begin{figure}[!h]
\centering
\begin{tikzpicture}[>=stealth, node distance=1.5cm]
    \tikzstyle{format} = [draw, very thick, ellipse,
                          minimum size=0.2cm, inner sep=1pt]
    \tikzstyle{unobs} = [draw, very thick, red, ellipse,
                         minimum size=0.2cm, inner sep=1pt]
    \tikzstyle{square} = [draw, very thick, rectangle,
                         minimum size=0.2cm, inner sep=2pt]

    \begin{scope}[xshift=4.5cm]
        \path[->, very thick]
		node[format] (A) {$A_0$}
		node[ above of=A] (C) {} 
		node[format, right of=A] (L1) {$C_1$}
		node[format, red, above of=L1] (U) {${U_0}$}

		node[format, right of=L1] (A1) {$A_1$}
		node[format, red, above of=A1] (U1) {$U_1$}
		node[format, right of=A1] (Y) {$Y$}

		node[format, below of=A] (Z) {$Z_0$}
		node[format, below of=L1] (W) {$W_0$}
        
		node[format, below of=A1] (Z1) {$Z_1$}
		node[format, below of=Y] (W1) {$W_1$}


		(U) edge[red] (A)
		(U) edge[red] (L1)
		(U) edge[red, bend left=0] (Y)
		(U) edge[red, bend left=90] (W1)

		(U) edge[red, bend left=25] (W)
		(U) edge[red] (Z)
		(U) edge[red] (Y)
		(U) edge[red] (U1)
		(U) edge[red] (A1)

		(Z) edge[blue] (A)
		(Z) edge[blue, bend right=25] (Z1)
		(Z) edge[blue, bend right=0] (A1)

		(W) edge[blue] (L1)
		(W) edge[blue, bend right=25] (W1)
		(W) edge[blue, bend right=15] (Y)

		(A) edge[blue] (L1)
		(A) edge[blue, bend right=15] (Z1)
		(A) edge[blue, bend left=20] (A1)
		(A) edge[blue, bend right] (Y)
		(A) edge[blue] (U1)

		(U1) edge[red] (A1)
		(U1) edge[red] (Y)
		(U1) edge[red] (W1)
		(U1) edge[red, bend left=25] (Z1)

		(A1) edge[blue] (Y)

		(Z1) edge[blue] (A1)

		(W1) edge[blue] (Y)

		(L1) edge[blue, bend right=0] (A1)
		(L1) edge[blue, bend right=20] (Y)
		(L1) edge[blue] (Z1)
		(L1) edge[blue] (W1)

        node[below of=W, yshift=0.8cm, xshift=0.75cm] (l) {$(a)$}
        ;
    \end{scope}
    
    \begin{scope}[xshift=10.5cm, yshift=-0.0cm]
        \path[->, very thick]

		node[format] (A) {$A_0$}
		node[ above of=A] (C) {} 
		node[format, right of=A] (L1) {$C_1$}
		node[format, red, above of=L1] (U) {${U_0}$}

		node[square, right of=L1] (A1) {$a_1$}
		node[format, red, above of=A1] (U1) {$U_1$}
		node[format, right of=A1] (Y) {$Y$}

		node[format, below of=A] (Z) {$Z_0$}
		node[format, red, below of=L1] (W) {$W_0$}
        
		node[format, below of=A1] (Z1) {$Z_1$}
		node[format, red, below of=Y] (W1) {$W_1$}


		(U) edge[red] (A)
		(U) edge[red] (L1)
		(U) edge[red, bend left=0] (Y)
		(U) edge[red, bend left=90] (W1)

		(U) edge[red, bend left=25] (W)
		(U) edge[red] (Z)
		(U) edge[red] (Y)
		(U) edge[red] (U1)

		(Z) edge[blue] (A)
		(Z) edge[blue, bend right=25] (Z1)

		(W) edge[red] (L1)
		(W) edge[red, bend right=25] (W1)
		(W) edge[red, bend right=15] (Y)

		(A) edge[blue] (L1)
		(A) edge[blue, bend right=15] (Z1)
		(A) edge[blue, bend right] (Y)
		(A) edge[blue] (U1)

		(U1) edge[red] (Y)
		(U1) edge[red] (W1)
		(U1) edge[red, bend left=25] (Z1)

		(A1) edge[blue] (Y)


		(W1) edge[red] (Y)

		(L1) edge[blue, bend right=20] (Y)
		(L1) edge[blue] (Z1)
		(L1) edge[blue] (W1)

        node[below of=W, yshift=0.8cm, xshift=0.75cm] (l) {$(b)$}
        ;
    \end{scope}
    
\end{tikzpicture}
\caption{
(a) A causal diagram corresponding to the proximal generalization of the sequentially ignorable model.
(b ) A causal diagram obtained from (a) after the treatment $A_1$ is fixed using proximal causal inference with the 
{outcome-inducing} proxy $W_1$.
}
\label{fig:g-comp}
\end{figure}

\section{An Important Special Case: The Proximal G-Computation Algorithm}
\label{appendix:g-comp}

Here we consider an important special case of the proximal ID algorithm that arises in longitudinal studies, where treatments are administered sequentially over time.
If the policy assigning treatments at each time point in such a study was based only on the past observed variables, the resulting causal model is the well known \emph{sequentially ignorable model} where identification is achieved by means of the g-computation algorithm \citep{robins86new}.  An example of such a model is shown in Fig.~\ref{fig:backdoor-2} (b), where we have
{\small
\begin{align}
p(Y(a_0, a_1) = y) = p(y \mid \text{do}(a_0, a_1)) = \sum_{c_0, c_1} p(y \mid a_0, c_0, a_1, c_1) p(c_1 \mid a_0, c_0) p(c_0).
\label{eqn:g-comp}
\end{align}
}
If the policy assigning treatments at each time point also depends on unobserved variables, the distribution $p(Y(a_0, a_1))$ is not identified nonparametrically.  However, provided the appropriate 
{outcome-inducing} and control proxies exist, we can exploit the proximal ID algorithm to obtain identification.  Consider the model shown in Fig.~\ref{fig:g-comp} (a), where we are interested in identifying $p(Y(a_0, a_1))${, and the vector of baseline covariates $C_0$, which is assumed to cause all other variables, is suppressed from the figure for legibility.
}

{
Let $\vec{z} = (z_0, z_1), \vec{c} = (c_0, c_1), \vec{a} = (a_0,a_1), \vec{w} = (w_0,w_1), \vec{u} = (u_0, u_1)$.
}
Identification proceeds in two stages, with both reusing and inductive margins initialized as $p(y, \vec{c}, \vec{a}, \vec{z}, \vec{w})$.
For the first stage, we make the following assumptions, implied by the graphical causal model.

\begin{assumption}[sequential proxy independence \#1]
\label{a:s-proxy-g-in}
{\small
\begin{align}
Y(a_1) &\ci Z_0, Z_1 \mid U_0, U_1, A_0, C_0, C_1{,A_1} \label{eqn:g-c-1}\\
W_1, W_0 &\ci A_1, Z_0, Z_1 \mid U_0, U_1, A_0, C_0, C_1 \label{eqn:g-c-2}
\end{align}
}
\end{assumption}

\begin{assumption}[sequential ignorability \#1]
\label{a:s-proxy-g-ig}
{\small
\begin{align}
Y(a_1) &\ci A_1 \mid 
U_0, U_1, A_0, C_0, C_1, \label{eqn:g-c-3}
\end{align}
}
\end{assumption}

{In addition, we assume the following bridge function and completeness conditions.}

\begin{assumption}[sequential outcome bridge function \#1]
\begin{itemize}
\item[]
\end{itemize}
{There exists a bridge function $b_{A_1}(w_1, y, a_1, z_0, a_0, c_0, w_0, c_1)$ such that the following equality holds:}
{\small
\begin{align}
p(y \mid a_1, z_0, a_0, c_0, w_0, c_1, z_1) = \sum_{w_1,w_0} b_{A_1}(w_1, y, a_1, z_0, a_0, c_0, w_0, c_1) p(w_1,w_0 \mid a_1, z_0, a_0, c_0, c_1, z_1)
\label{eqn:g-c-4}
\end{align}
}
\end{assumption}
 
\begin{assumption}[sequential completeness \#1]
\begin{itemize}
\item[]
\end{itemize}
{For any random function $v(U_1)$,}
{\small
\begin{align}
\sum_{U_1} p(v(U_1) \mid a_1, z_0, a_0, c_0, w_0, c_1, z_1) = 0 \text{ for all }a_1, z_0, a_0, c_0, w_0, c_1, z_1\text{ if and only if }v(U_1) = 0 \label{eqn:g-c-5}.
\end{align}
}
\end{assumption}

{We first obtain the following:
{\small
\begin{align}
(\ref{eqn:g-c-1}) &\underbrace{\Rightarrow}_{\text{consistency}} Y \ci Z_0, Z_1 \mid A_1{=a_1}, U_0, U_1, A_0, C_0, C_1 \label{eqn:g-c-6}\\
\notag
(\ref{eqn:g-c-1}) + (\ref{eqn:g-c-3}) &\underbrace{\Rightarrow}_{\text{contraction}} Y(a_1) \ci Z_0, Z_1,A_1 \mid U_0, U_1, A_0, C_0, C_1\\
&\underbrace{\Rightarrow}_{\text{weak union}}  Y(a_1) \ci A_1 \mid U_0, U_1, A_0, C_0, C_1,Z_0, Z_1 \label{eqn:g-c-ig-z}\\
(\ref{eqn:g-c-2}) &\underbrace{\Rightarrow}_{\text{weak union}} W_1, W_0 \ci Z_0, Z_1 \mid U_0, U_1, A_0, C_0, C_1, A_1 \label{eqn:g-c-w}\\
(\ref{eqn:g-c-2}) &\underbrace{\Rightarrow}_{\text{weak union}} W_1, W_0 \ci A_1 \mid U_0, U_1, A_0, C_0, C_1, Z_0, Z_1 \label{eqn:g-c-w-2}
\end{align}
}
}


{Writing $b_{A_1}(w_1, y, a_1, z_0, a_0, c_0, w_0, c_1)$ as $b_{A_1}$ for conciseness,} we then have:
{\small
\begin{align*}
p(y \mid \vec{a}, \vec{z}, \vec{c}) &= \sum_{\vec{w}} b_{A_1}
	p(\vec{w} \mid \vec{a}, \vec{z}, \vec{c}) \Rightarrow (\text{by }(\ref{eqn:g-c-6}),(\ref{eqn:g-c-w})) \\
\sum_{\vec{u}}
p(y \mid \vec{a}, \vec{c}, \vec{u}) p(\vec{u} \mid \vec{a}, \vec{z}, \vec{c}) &=
\sum_{\vec{w} } b_A \sum_{\vec{u} } p(\vec{w} \mid \vec{a}, \vec{c}, \vec{u}) p(\vec{u} \mid \vec{a}, \vec{z}, \vec{c}) \Rightarrow
(\text{by }(\ref{eqn:g-c-5}))\\
p(y \mid \vec{a}, \vec{c}, \vec{u}) &= 
\sum_{\vec{w} } b_A p(\vec{w} \mid \vec{a}, \vec{c}, \vec{u}) \Rightarrow (\text{by }(\ref{eqn:g-c-6}),(\ref{eqn:g-c-w}),(\ref{eqn:g-c-w-2}))\\
\sum_{\vec{u}}
p(y \mid \vec{a}, \vec{z}, \vec{c}, \vec{u}) p(\vec{u}, a_0, \vec{z}, \vec{c}) &=
\sum_{\vec{u}} \sum_{\vec{w} } b_A p(\vec{w} \mid a_0, \vec{z}, \vec{c}, \vec{u}) p(\vec{u}, a_0, \vec{z}, \vec{c}) \Rightarrow (\text{by }(\ref{eqn:g-c-ig-z}))\\
\sum_{\vec{u}}
\frac{
p(y, \vec{a}, \vec{z}, \vec{c}, \vec{u})
}{
p(a_1 \mid \vec{u}, \vec{c}, \vec{z}, a_0)
}
&=
p(y, a_0, \vec{z}, \vec{c} \mid \text{do}(a_1)) = \sum_{\vec{w}} b_A p(\vec{w}, a_0, \vec{z}, \vec{c}).
\end{align*}
}
This results in the graph in Fig.~\ref{fig:g-comp} (b), and identification of the 
{inductive} margin $p(y, a_0, \vec{z}, \vec{c} \mid \text{do}(a_1))$.  Since $A_1$ is not fixable in
${\cal G}(\vec{C}, \vec{A}, \vec{Z}, \vec{W}, Y)$, we obtain the reusing margin by marginalizing out descendants of $A_1$, yielding
$p(\vec{c}, a_0, \vec{z}, \vec{w})$.

Since $Z_1$ has no descendants once $A_1$ is fixed, {and isn't required for any subsequent operation}, 
{$Z_1$} may be safely marginalized out from both margins, yielding:
{\small
\begin{align*}
p(y, a_0,z_0, \vec{c} \mid \text{do}(a_1)) &= \sum_{\vec{w}} b_A p(\vec{w}, a_0, z_0, \vec{c})\\
p(\vec{c}, a_0, z_0, \vec{w}) &= \sum_{z_1} p(\vec{c}, a_0, \vec{z}, \vec{w}).
\end{align*}
}
{This marginalization corresponds to the usual fixing operation described in \S \ref{sssec:fixing}.}

{To continue the induction,}
we employ the following assumptions for the second stage, which consist of independences implied by the causal model, a bridge integral equation {assumption}, and a completeness condition.

\begin{assumption}[sequential proxy independence \#2]
\label{a:s-proxy-g-in-2}
{\small
\begin{align}
Y(a_1, a_0), C_1(a_1, a_0) &\ci Z_0 \mid U_0, C_0{,A_0} \label{eqn:g-c2-1}\\
W_0 &\ci A_0, Z_0 \mid U_0, C_0 \label{eqn:g-c2-2}
\end{align}
}
\end{assumption}

\begin{assumption}[sequential ignorability \#2]
\label{a:s-proxy-g-ig-2}
{\small
\begin{align}
Y(a_1, a_0), C_1(a_1, a_0) &\ci A_0 \mid 
U_0, C_0 \label{eqn:g-c2-3}
\end{align}
}
\end{assumption}

\begin{assumption}[sequential outcome bridge function \#2]
\begin{itemize}
\item[]
\end{itemize}
{There exists a bridge function $b_{A_0}(w_0, y, c_1, a_0, c_0, a_1)$ such that the following equality holds:}
{\small
\begin{align}
p(y, c_1 \mid a_0, z_0, c_0, \text{do}(a_1)) &= \sum_{w_0} b_{A_0}(w_0, y, c_1, a_0, c_0, a_1) p(w_0 \mid a_0, z_0, c_0)
\label{eqn:g-c2-4}
\end{align}
}
\end{assumption}

\begin{assumption}[sequential completeness \#2]
\begin{itemize}
\item[]
\end{itemize}
{For any random function $v(U_0)$,}
{\small
\begin{align}
\sum_{U_0} p(v(U_0) \mid a_0, z_0, c_0, \text{do}(a_1)) &= 0 \text{ for all }z_0, a_0, c_0, a_1\text{ if and only if }v(U_0) = 0 \label{eqn:g-c2-5},
\end{align}
}
\end{assumption}
where the left hand side of (\ref{eqn:g-c2-4}) is a function of the inductive margin obtained from the first stage, and the right hand side of (\ref{eqn:g-c2-4}) is a function of the reusing margin obtained from the first stage.

{
We then obtain the following:
{\small
\begin{align}
(\ref{eqn:g-c2-1}) &\underbrace{\Rightarrow}_{\text{consistency}} Y(a_1), C_1(a_1) \ci Z_0 \mid U_0, C_0,A_0=a_0 \label{eqn:g-c2-6}\\
\notag
(\ref{eqn:g-c2-1}) +
(\ref{eqn:g-c2-3}) &\underbrace{\Rightarrow}_{\text{contraction}} Y(a_1, a_0), C_1(a_1, a_0) \ci Z_0,A_0 \mid U_0, C_0 \\
& \underbrace{\Rightarrow}_{\text{weak union}} Y(a_1, a_0), C_1(a_1, a_0) \ci A_0 \mid U_0, C_0,Z_0 \label{eqn:g-c2-ig-z}\\
(\ref{eqn:g-c2-2})
&\underbrace{\Rightarrow}_{\text{weak union}}
W_0 \ci Z_0 \mid U_0, C_0,A_0 \label{eqn:g-c2-w}\\
(\ref{eqn:g-c2-2})
&\underbrace{\Rightarrow}_{\text{weak union}}
W_0 \ci A_0 \mid U_0, C_0,Z_0 \label{eqn:g-c2-w-2}
\end{align}
}
}

We 
{next} obtain the following derivation for the second stage, {where $b_{A_0}$ denotes $b_{A_0}(w_0, y, c_1, a_0, c_0, a_1)$ for conciseness}:
{
\small
\begin{align*}
p(y, c_1 \mid a_0, z_0, c_0, \text{do}(a_1)) &= \sum_{w_0} b_{A_0} p(w_0 \mid a_0, z_0, c_0) \Rightarrow (\text{by }(\ref{eqn:g-c2-6}),(\ref{eqn:g-c2-w}))\\
\sum_{u_0} p(y, c_1 \mid u_0, a_0, c_0, \text{do}(a_1)) p(u_0 \mid a_0, z_0, c_0) 
&=
\sum_{w_0} b_{A_0}\! \sum_{u_0} p(w_0 | a_0, u_0, c_0) p(u_0 | a_0, z_0, c_0) \Rightarrow\\
& \hspace{4.0cm} (\text{by }(\ref{eqn:g-c2-5}))\\
p(y, c_1 \mid u_0, a_0, c_0, \text{do}(a_1)) &= \sum_{w_0} b_{A_0} p(w_0 \mid a_0, u_0, c_0) \Rightarrow\\
& \hspace{4.0cm} (\text{by }(\ref{eqn:g-c2-6}),(\ref{eqn:g-c2-w}),(\ref{eqn:g-c2-w-2}))\\
\sum_{u_0} p(y, c_1 \mid u_0, a_0, z_0, c_0, \text{do}(a_1)) p(u_0 \mid c_0, z_0, \text{do}(a_1))
&=
\sum_{u_0} \sum_{w_0} b_{A_0} p(w_0 \mid u_0, z_0, c_0) p(u_0 \mid c_0, z_0) \Rightarrow\\
& \hspace{4.0cm} (\text{by }(\ref{eqn:g-c2-ig-z}))\\
\sum_{u_0}
\frac{
p(y, c_1, u_0, a_0, z_0, c_0 \mid \text{do}(a_1))
}{
p(a_0 \mid c_0, z_0, u_0, \text{do}(a_1))
} &= p(y, c_1, z_0, c_0 \mid \text{do}(a_1, a_0))\\
&= \sum_{w_0} b_{A_0} p(w_0, z_0, c_0).
\end{align*}
}
Together, the above two derivations imply the following identification result:
{\small
\begin{align*}
p(y \mid \text{do}(a_1, a_0)) &= \sum_{w_0,c_1,c_0} b_{A_0} p(w_0, c_0), \text{ where }b_{A_0}\text{ is solved for via}\\
p(y, c_1 \mid a_0, z_0, c_0, \text{do}(a_1)) &= \sum_{w_0} b_{A_0}(w_0, y, c_1, a_0, c_0, a_1) p(w_0 \mid a_0, z_0, c_0), \text{ where}\\
p(y, c_1 \mid a_0, z_0, c_0, \text{do}(a_1)) &=
\frac{
\sum_{\vec{w}} b_A p(\vec{w}, a_0, z_0, \vec{c})
}{
\sum_{y, c_1,\vec{w}} b_A p(\vec{w}, a_0, z_0, \vec{c})
},\text{ since}\\
p(y, a_0,z_0, \vec{c} \mid \text{do}(a_1)) &= \sum_{\vec{w}} b_A p(\vec{w}, a_0, z_0, \vec{c}).
\end{align*}
}

This section provides a graphical description of the proximal g-computation algorithm first described in \citep{tchetgen2020introduction}.  We note a few differences between the algorithm in this section, and the one in \citep{tchetgen2020introduction}.  First, the target of inference in \citep{tchetgen2020introduction} was a counterfactual expectation, and thus all derivations employed integral equations on the expectation scale.  Second, control proxies $Z_1$ and $Z_0$, while used in the same way in both versions of the algorithm, were immediately removed in \citep{tchetgen2020introduction} upon use.  In our version of the algorithm, they are kept around after their use, and are subsequently removed by means of the fixing operation.  This step was made explicit to emphasize the connection between the proximal ID algorithm, and the (classical) ID algorithm.

Finally, the assumptions employed in the second stage of the derivation in \citep{tchetgen2020introduction} were formulated on potential outcomes derived from $Y_1$, rather than on the joint potential outcomes derived from $Y_1$ and $C_1$ as is done here.   This is possible because the marginalization over $C_1$ can be performed earlier in the algorithm while retaining validity of the overall derivation.  The advantage of formulating assumptions only on $Y_1$ becomes apparent in high dimensional applications, where the bridge integral equation becomes much easier to formulate if fewer variables are involved.  In this paper, the marginalization over $C_1$ is performed at the end of the algorithm, to better illustrate the connection with the operation of the (classical) ID algorithm.  A general method for performing summations in a way that results in lower dimensional integral equations is closely related to methods for efficient marginalization in graphical models via \emph{sum product algorithms}, and is left to future work.

\bibliography{references}

\end{document}